\documentclass[12pt]{article}

\def\bphi{\mbox{\boldmath $\phi$}}

\def\balpha{\mbox{\boldmath $\alpha$}}

\def\m{\mbox{\bf m}}

\input{amssymb.sty}

\def\tr{\mathop{\rm tr}\nolimits}
\def\Tr{\mathop{\rm Tr}\nolimits}
\def\frac#1#2{{{#1}\over{#2}}}
\def\tfrac#1#2{{\textstyle{{#1}\over{#2}}}}

\def\half{\tfrac{1}{2}}
\def\third{\tfrac{1}{3}}

\begin{document}

\begin{titlepage}

\baselineskip 24pt

\begin{center}

{\Large {\bf A Model Behind the Standard Model}}

\vspace{.5cm}

\baselineskip 14pt

{\large CHAN Hong-Mo}\\
h.m.chan\,@\,rl.ac.uk \\
{\it Rutherford Appleton Laboratory,\\
  Chilton, Didcot, Oxon, OX11 0QX, United Kingdom}\\
\vspace{.2cm}
{\large TSOU Sheung Tsun}\\
tsou\,@\,maths.ox.ac.uk\\
{\it Mathematical Institute, University of Oxford,\\
  24-29 St. Giles', Oxford, OX1 3LB, United Kingdom}

\end{center}

\vspace{.3cm}

\begin{abstract}

In spite of its many successes, the Standard Model makes many empirical 
assumptions in the Higgs and fermion sectors for which a deeper theoretical 
basis is sought.  Starting from the usual gauge symmetry $u(1) \times 
su(2) \times su(3)$ plus the 3 assumptions: (A) scalar fields as vielbeins
in internal symmetry space \cite{framevec}, (B) the ``confinement picture''
of symmetry breaking \cite{tHooft,Banovici}, (C) generations as ``dual'' 
to colour \cite{genmixdsm}, we are led to a scheme which offers: (I) a 
geometrical significance to scalar fields, (II) a theoretical criterion on
what scalar fields are to be introduced, (III) a partial explanation of why
$su(2)$ appears broken while $su(3)$ confines, (IV) baryon-lepton number
(B - L) conservation, (V) the standard electroweak structure, (VI) a 3-valued
generation index for leptons and quarks, and (VII) a dynamical system with all 
the essential features of an earlier phenomenological model \cite{genmixdsm}
which gave a good description of the known mass and mixing patterns of 
quarks and leptons including neutrino oscillations.  There are other 
implications the consistency of which with experiment, however, has not yet 
been systematically explored.  A possible outcome is a whole new
branch of particle spectroscopy from 
$su(2)$ confinement, potentially as rich in details as that of 
hadrons from colour 
confinement, which will be accessible to experiment at high energy.

\end{abstract}

\end{titlepage}

\clearpage

\baselineskip 14pt

\setcounter{equation}{0}

\section{Introduction}

Despite its many successes, the Standard Model as it now stands gives one
the impression of being but the consequence of a deeper theory yet to be
divined.  The many input parameters on which it depends together with the
intricate structural details which have to be built into it are a little 
beyond what one would expect of a truly fundamental theory.  At a deeper 
level, one can wonder of course why nature should opt for a gauge symmetry 
of $u(1) \times su(2) \times su(3)$ and not some other symmetry, or why 
there should be a gauge structure in the first place.  But even if one takes 
this particular gauge structure for granted, one finds that one has still
to add quite a number of ingredients mostly connected to the Higgs and 
fermion sectors which are seemingly extraneous to the gauge hypothesis.  For 
example, to break the electroweak $su(2)$ symmetry so as to fit the picture 
obtained from experiment, one introduces in the Standard Model an $su(2)$ 
doublet of scalar fields, whose significance in the original gauge framework 
is a little obscure, which is a pity, for the bosonic sector of the framework 
is otherwise so geometrical.  Furthermore, we have not been able to explain 
why of the two nonabelian symmetries, the electroweak symmetry $su(2)$ should 
be broken, while the colour symmetry $su(3)$ is not.  In the fermion sector 
also, nature tells us that we have to introduce 3 generations for each of 
the 4 fermion species: $U$-type quark, $D$-type quark, charged leptons and 
neutrinos, and that we have to take left-handed fermions as $su(2)$ doublets 
and right-handed fermions singlets, without us being able to fathom why she 
would want us to do so.  Indeed, it is this our inability to answer the 
above questions in the Higgs and fermion sectors which forces on us the bulk 
of the twenty-odd independent parameters that have to be fed into the Standard 
Model.  Furthermore, at the secondary, detail level, there are the no less
intriguing questions why the observed fermion masses should be hierarchical,
with values differing from generation to generation by orders of magnitude, 
and why the mixing angles between up and down fermion states be so different 
from case to case, ranging from order $10^{-3}$ for $V_{ub}$, say, to order 
unity for the oscillation of atmospheric and solar neutrino, which questions 
are also left unanswered in the present Standard Model. Any understanding of 
them will thus be not only aesthetically satisfying but also of practical 
value, however incomplete the understanding we may be able to achieve at the 
present stage.

To attempt answering these questions, one can try extending the theory by
enlarging the symmetry as in grand unified and supersymmetric theories, or
by increasing the number of space-time dimensions as in Kaluza-Klein type
theories, and/or the dimensions of the fundamental object as well, as in
strings and branes.  The tendency of such attempts, however, is to increase
the number of unknown parameters rather than to reduce it.  An alternative 
is to forego for the moment the very ambitious vision of the above attempts, 
attractive though it may be, opting instead for economy, and try to answer
the questions posed all within the framework of the Standard Model.

It is in this latter spirit that we have attempted here to construct, based 
on some simple and hopefully reasonable assumptions, a new ``protogenic'' 
model which will give the Standard Model as the result.  The 
attempt builds on three ideas which have been suggested earlier in slightly 
different contexts.  The first (A) is the proposal that frame vectors in the 
gauge symmetry space be promoted to fields so as eventually to play the 
role of Higgs fields, thus giving the latter a geometrical significance which 
they at present lack \cite{physcons,framevec}.  The second (B) is an old idea 
of 't~Hooft \cite{tHooft} and of Banks and Rabinovici \cite{Banovici}, 
re-emphasized 
more recently by 't~Hooft \cite{tHooft1}, that symmetry breaking in certain 
circumstances may be re-interpreted as the consequence of a confining theory.  
The third (C) is the suggestion that the generation symmetry is in some sense 
dual to colour $su(3)$ symmetry \cite{physcons,genmixdsm}, so that one 
obtains automatically 3 and only 3 generations of fermions.  These three 
ideas, as we hope to show, can bring us quite close to constructing a 
protogenic model behind the Standard Model as desired.  Besides, they can 
each lead to quite revolutionary changes in our general concepts, which we 
wish first now to outline.

(A) The idea that scalar fields appearing in gauge theories may have
the geometrical significance of frame vectors in internal symmetry space
is analogous to the familiar concept of vierbeins being introduced as 
dynamical variables in gravitation theory.  It means that these scalar 
fields, having a specific geometrical function to discharge, are to be 
regarded as an integral part of the gauge structure, and not to be
introduced or discarded at will to fit our interpretation of data or some 
other prejudice.  Their existence and properties are to be determined by 
the gauge symmetries in the theory leaving us little room for choice.  
Whether they can function as Higgs fields to break the appropriate symmetries 
and give a realistic model of nature has yet to be seen, but if they do, 
then the Higgs mechanism as normally conceived say in electroweak theory 
will cease to be an input assumption and become just a consequence of an 
all-enveloping gauge concept. 

(B) It was shown by 't~Hooft and Banks and Rabinovici already in the
late seventies that the electroweak theory which is usually conceived as a 
theory with a spontaneously broken  $su(2)$ gauge symmetry can equivalently
be considered as a theory where the $su(2)$ gauge symmetry confines; what
is actually broken is  a global $su(2)$ symmetry which is associated with 
but is not the same as the original local gauge symmetry.  If we accept the 
latter interpretation, then the difference between $su(2)$ and $su(3)$ in 
the Standard Model is not any more a matter of status, i.e. whether 
spontaneously broken or confining, but just a matter of degree, i.e. how 
deeply the symmetry is confined so as to be accessible or not to probing 
by present experiment.  If that is the case then it is a matter to be 
understood in terms of the dynamics and is not to be regarded as part of 
the empirical input.

(C) The idea that fermion generations may in some sense be dual to
colour introduces automatically into the theory exactly 3 generations of
fermions, the existence of which therefore need no longer to be assumed.  
Besides, the supposition brings with it its own dynamical logic so that 
the parameters which characterise the 3 fermion generations can now appear 
as dynamical consequences and again be removed from the Standard Model as 
inputs from experiment, thus drastically reducing the number of empirical 
parameters.  Indeed, previously, with a phenomenological model we called
the Dualized Standard Model (DSM) built roughly along these lines, we were 
already able to reproduce correctly most of the fermion mass ratios and 
mixing parameters \cite{phenodsm,genmixdsm}.  

One sees therefore these ideas can in principle go a long way towards our
goal of reducing the degree of arbitrariness in the formulation and serve 
as a basis for the construction of a protogenic version to the Standard 
Model. However, whether they will be able, on being put to practice, to 
produce a model approximating nature is a question which can be answered 
only by carrying out the program explicitly.  The following is an outline 
of the logic followed and conclusions obtained in our attempt, which are 
to be detailed in the succeeding sections.

The idea (A) as developed in \cite{framevec} specifies to a large extent 
what scalar fields corresponding to frame vectors (called framons 
henceforth) are to be introduced for a given gauge symmetry.  When 
supplemented by an appeal to economy, it leads for the gauge symmetry 
$u(1) \times su(2) \times su(3)$ to 2 sets of framons: a ``weak'' framon
transforming as a doublet of $su(2)$ and 3 ``strong'' framons transforming
as triplets of $su(3)$.  As frame vectors, they give the orientation of the 
local frames with respect to some global reference frames, and hence carry 
with them indices referring to the global frames as well.  Since physics
should not depend on the choice of reference frames, it follows that one
has invariance not only under the original local gauge symmetries but also
under the global symmetries $\tilde{u}(1) \times \widetilde{su}(2) \times
\widetilde{su}(3)$.  This doubled invariance puts a stringent condition
on the framon action, in particular on the framon self-interaction potential
which, up to 4th order for renormalizability, is found to have a unique form
depending on 7 real parameters.  The part of the potential depending only 
on the ``weak'' framon is identical in form to the standard electroweak 
potential, that only on the ``strong'' framons is similar in form to the 
scalar potential in DSM, but there are additional terms linking the ``weak'' 
and ``strong'' sectors.  The vacuum for this potential is degenerate and 
depends, by virtue of the linkage terms, on the orientation of the ``weak'' 
framon in $\widetilde{su}(3)$ space.

Next, in the confinement picture of \cite{tHooft} and \cite{Banovici} adopted
in (B), all the local symmetries $u(1) \times su(2) \times su(3)$ remain
exact, and both nonabelian symmetries confine, with the confinement by
$su(2)$ much deeper than that by (colour) $su(3)$.  What are broken are the
global symmetries $\widetilde{su}(2)$ and $\widetilde{su}(3)$, the first
by the $u(1)$ gauge interaction, the second by the framon potential above
via the linkage to the ``weak'' sector.  Under the present experimental
regime where one already probes routinely into hadronic structures (call 
this the standard model scenario), one needs consider only $su(2)$ as 
confining.  In that case, only $su(2)$ singlets are observable and still
appear as elementary.  In particular, as in the electroweak theory treated 
in \cite{tHooft} and \cite{Banovici}, the Higgs boson appears as the 
$s$-wave bound state by $su(2)$ confinement of the ``weak'' framon with 
its conjugate, while the vector bosons $W^{\pm}, Z-\gamma$ appear as 
$p$-wave bound states, and one recovers the standard electroweak theory 
as the result.

Introducing next as fundamental fermion fields the simplest
representations, say $\psi(1,1)$,
$\psi(1,3)$, $\psi(2,1)$, $\psi(2,3)$, where the first number denotes the 
$su(2)$ and the second the $su(3)$ representation, one can form bound states 
from the last 2 with the ``weak'' framon by $su(2)$ confinement, which, as 
in the electroweak theory treated in \cite{tHooft}, \cite{Banovici}, and 
\cite{framevec}, represent respectively the left-handed leptons and quarks.  
There is one difference, however, namely that here the ``weak'' framon carry 
indices referring to the global symmetries $\tilde{u}(1), \widetilde{su}(2)$, 
and $\widetilde{su}(3)$, and these are now transmitted to the fermion bound 
states, i.e the left-handed leptons and quarks.  The conserved $\tilde{u}(1)$ 
charge is found to be the baryon-lepton number, the $\widetilde{su}(2)$ 
symmetry broken by $u(1)$ is identified already in \cite{tHooft,Banovici}
and \cite{framevec} as up-down flavour, while $\widetilde{su}(3)$, broken 
in a rather special manner to be outlined in the next paragraph, is to play 
the role of generations, as proposed in (C).

The Yukawa coupling constructed to have the required invariance gives for 
the tree-level mass matrix for both leptons and quarks a factorized form:
\begin{equation}
m = m_T \balpha \balpha^{\dagger}
\label{massmatlq}
\end{equation}
where $m_T$ depends on the fermion species, but $\balpha$, a vector in
$\widetilde{su}(3)$ space coming from the ``weak'' framons, does not.  
The vector $\balpha$, however, is coupled to the ``strong'' framons via
the linkage terms in the framon potential.  An examination of the dynamics
in the strong sector to first perturbative order along the lines already
performed in \cite{ckm} for DSM shows that $\balpha$ will rotate in
generation space as the scale $\mu$ changes, of which rotation there is 
a fixed point at $\mu = 0$ and another at $\mu = \infty$, and $\balpha$
rotates away from the first fixed point towards the second as $\mu$
increases.  Now these properties of the fermion mass matrix are exactly
those that were found in earlier analyses \cite{phenodsm} to be essential 
and very likely sufficient to give a reasonable description of the fermion 
mass and mixing patterns observed in experiment.  It seems thus hopeful
that similar agreement with experiment as for DSM can be achieved here
although this can only be confirmed by an explicit calculation, which
is now being pursued.

 In view of these results, we venture to conclude that one does seem to
have gone some way towards what we called a protogenic model behind the
Standard Model.  There are several unanswered questions, and distinguishing
predictions to be tested, which we reserve for the concluding section,
including in particular a whole possible new field of particle 
spectroscopy to be opened up, perhaps at LHC, but if not, then when 
high enough experimental energies become available.

\setcounter{equation}{0}

\section{The Fundamental Boson Fields}

We start with a theory with gauge symmetry $u(1) \times su(2) \times
su(3)$.  Here and throughout this paper, we denote a gauge symmetry by 
its gauge (Lie) algebra whenever there is no necessity to specify which
among the locally isomorphic Lie groups is to be selected as the gauge 
group, so as to avoid questions of topology inessential for the problem
at hand.  By convention, lower case letters denote algebras and upper 
case letters groups.

We introduce first as usual for the 3 factors respectively the gauge
fields $A_\mu(x), B_\mu(x), C_\mu(x)$ with the well-known geometric
significance of connections in the corresponding principal bundles.

In addition, following the suggestion (A) in the Introduction, we shall
introduce as part of the gauge structure Lorentz scalar fields having
the geometric significance of frame vectors in internal symmetry space,
which we call here framons.  The concept that frame vectors or vielbeins 
can be introduced as dynamical variables is familiar already in gravity
theory.  It thus requires no great stretch of imagination to consider
having them as dynamical variables in gauge theories as well.  Indeed,
if gauge structures were to be obtained by compactification of higher 
dimensions as they are in certain string and Kaluza-Klein type theories, 
then it would seem natural, perhaps even necessary, to introduce, along
with the vierbeins of gravity, such framons for gauge theories also.

Let us first make clear what we mean by frame vectors and framons in each
of the 3 simple factors of the symmetry $u(1) \times su(2) \times su(3)$
of present interest, repeating briefly here some considerations in 
\cite{framevec} for completeness.  For the $su(N)$ symmetries, 
frame vectors can be 
taken as the column vectors $\bphi^{\tilde{a}}$ of the matrix:
\begin{equation}
\Phi = (\phi_a^{\tilde{a}}),
\label{Phi}
\end{equation}
which specifies the orientation of the local ($x$-dependent) $su(N)$ 
frame with respect to some global ($x$-independent) reference frame, 
where $a = 1,2,...,N$ which labels the rows refers to the local frame 
and $\tilde{a} = 1,2,...,N$ which labels the columns refers to the global 
reference frame.  By definition then, $\Phi$ transforms by a fundamental 
representation of the local $su(N)$ operating from the left, but by an 
anti-fundamental representation of a global $\widetilde{su}(N)$ operating 
from the right, where the latter represents the effect on $\Phi$ by a 
change in the global reference frame.

As a transformation matrix between 2 $su(N)$ frames (i.e. the local and
the global), $\Phi$ would satisfy, of course, the unitary constraints:
\begin{equation}
\Phi^{\dagger} \Phi = 1, \ \ \ {\rm det}(\Phi) = 1,
\label{Ucon}
\end{equation}
which means, in particular, that the frame vectors $\bphi^{\tilde{a}}$
would each be of unit length.  As framon fields, however, we would want 
the components of $\bphi^{\tilde{a}}$, in analogy to the vierbeins of 
gravity, to have the freedom to vary over all (in this case, complex) 
values, so that (\ref{Ucon}) can no longer be fully satisfied.  We could, 
of course, allow the components of $\Phi$ to vary independently over all 
complex values and ignore the constraint (\ref{Ucon}) altogether, thus 
introducing all the $N^2$ complex or $2 N^2$ real components of $\Phi$ 
as independent field variables.  But this would seem extravagant, since
these components need not all be independent.  In order to minimize, 
for the sake of economy, the total number of independent fields to be 
introduced, we propose instead that we retain as much of (\ref{Ucon}) 
as is consistent with the desired freedom and with the natural condition 
that all framons have the same physical dimension.  

For $su(2)$, we see that we can retain from (\ref{Ucon}) the condition:
\begin{equation}
\phi_r^{\tilde{2}} = - \epsilon_{rs} (\phi_s^{\tilde{1}})^*,
\label{su2ortho}
\end{equation}
namely $\bphi^{\tilde{1}}$ and $\bphi^{\tilde{2}}$ being orthogonal and 
having the same length, but still allow all their components the freedom 
to range over all complex values.  Besides, the relation being linear 
and homogeneous in $\bphi^{\tilde{1}}$ and $\bphi^{\tilde{2}}$ guarantees 
that they can remain of the same physical dimension.  This means that 
for $su(2)$ theory, we can eliminate via (\ref{su2ortho}) one of the 2 
framon fields in terms of the other and be left with only one independent
$su(2)$ doublet, say $\bphi = \bphi^{\tilde{1}}$, i.e. 2 complex or 4 
real components as independent field variables, halving the number of 
the original components in $\Phi$.  The same tactic, however, would 
not work for $su(3)$, nor indeed for any $su(N)$ with $N > 2$.  We could
of course also reduce the number of independent framons in $su(3)$ by 
insisting that the vector $\bphi^{\tilde{3}}$, say, be always orthogonal 
to $\bphi^{\tilde{1}}$ and $\bphi^{\tilde{2}}$ by retaining from 
(\ref{Ucon}) the condition:
\begin{equation}
\phi_r^{\tilde{3}} 
   = \epsilon_{rst} (\phi_s^{\tilde{1}})^* (\phi_t^{\tilde{2}})^*,
\label{su3ortho}
\end{equation}
but this relation is inhomogeneous, implying for $\bphi^{\tilde{3}}$ 
a different physical dimension from that of $\bphi^{\tilde{1}}$ and 
$\bphi^{\tilde{2}}$, a conclusion we cannot physically accept.  The 
best that we can do, it seems, is just to retain from (\ref{Ucon}) the
condition that the determinant be real.  When cast in the form:
\begin{equation}
{\rm det} (\Phi) = ({\rm det} (\Phi))^*,
\label{detreal}
\end{equation}
the condition is homogeneous, thus not suffering from the objection
against (\ref{su3ortho}) above, and multilinear, thus allowing the 
independent variables to attain all values as demanded.  This removes
only one real component from the original 18 in $\Phi$ but seems already 
the best economy that can be achieved in $su(3)$.  As will be seen, this 
difference between $su(2)$ and $su(3)$, due just to their different 
structures, will play a significant role in this scheme in reproducing 
the very different physics arising from the 2 nonabelian symmetries in 
the Standard Model.

Next, let us repeat the above considerations for the remaining $u(1)$ 
factor.  Here orientation means just a phase, and relative orientation
just a phase difference.  Hence, the analogue of $\Phi$ above for the
$su(N)$ factors is here just a phase factor of the form:
\begin{equation}
\Phi = \exp i g_1 (\alpha - \tilde{\alpha}),
\label{Phiu1}
\end{equation}
with $\alpha$ $x$-dependent, transforming under the local $u(1)$ but 
$\tilde{\alpha}$ $x$-independent, transforming under the global 
$\tilde{u}(1)$ transformations.  The framon field is then just a complex
scalar field with its phase as in (\ref{Phiu1}) above.  

Notice that the framon fields so introduced, simply by virtue of their 
assigned geometric significance as frame vectors, have a special property
not shared by the gauge fields $A_\mu, B_\mu$ and $C_\mu$ introduced
before nor by the fermion fields yet to be introduced.  Namely, apart
from transforming under the local (gauge) transformation as other fields 
do, they carry in addition a global index ($\tilde{a}$ for the nonabelian
symmetries and $\tilde{\alpha}$ for $u(1)$) which transforms under global
changes of the reference frame.  Since physics should not be affected by 
the choice of reference frames, the action for framons should be invariant 
also under such global changes.  As will be seen, this helps to restrict 
the form of the action and leads automatically to additional conserved 
quantum numbers, which are assigned in the present scheme to such otherwise 
unexplained quantities as the baryon-lepton number and the (fermion) 
generation index. 

Having now specified what is meant by framons for each of the simple factors 
in the symmetry $u(1) \times su(2) \times su(3)$ of actual interest, we turn 
now to the physical problem itself.  Let us first consider the electroweak 
sector characterized by the local symmetry $u(1) \times su(2)$ and ask here
what framons should be introduced.  We start again with a matrix giving 
the orientation of the local frame with respect to some global reference 
frame but now for the symmetry $u(1) \times su(2)$.  The columns of this
matrix are what we call the frame vectors, which, as for the previous 
cases, are to be representations of the local symmetry $u(1) \times su(2)$.
Now, from the representations of the simple factor symmetries $u(1)$ and 
$su(2)$, there are 2 ways to build a representation of $u(1) \times su(2)$, 
i.e. taking either the sum or the product.  Supposing we appeal again to 
economy and opt for the choice requiring the smallest number of independent 
framon fields, we would choose the product, since $1 \times 2 < 1 + 2$.
In other words, the frame vectors would again be $su(2)$ doublets, but now
carrying each a $u(1)$ charge.  Furthermore, by virtue of (\ref{su2ortho}), 
we would eliminate one of these, leaving just one charged doublet, say, 
$\bphi = \bphi^{\tilde{1}}$ as the only framon field. 

We have yet to specify what $u(1)$ charge this field $\bphi$ should carry.
This depends not just on the gauge algebra which is $u(1) \times su(2)$, 
but on the choice of the gauge group.  Now there are 3 locally isomorphic
groups with this same algebra, namely: $U(1) \times SU(2)$, $U(2) = (U(1) 
\times SU(2))/Z_2$, and $U(1) \times SO(3)$, of which 3 we can discard 
immediately the last since it does not contain the $su(2)$ doublet as a 
representation.  Of the remaining 2, the first, $U(1) \times SU(2)$, 
double-covers the second $U(2)$.  Suppose we appeal again to minimality 
and choose the the smaller group $U(2)$, we end up with a half-integral 
charge, say $\pm g_1/2$, for the single doublet framon field $\bphi$.  
We notice that this is identical in properties to the Higgs field in 
standard electroweak theory.  Indeed, starting with the framon concept, 
insisting on global $\tilde{u}(1) \times \widetilde{su}(2)$ invariance
in addition to the usual local gauge $u(1) \times su(2)$ invariance in 
constructing the action, and then following the above prescription based 
on minimality, one would be led uniquely to the standard electroweak 
theory as the result.  A demonstration of this is given in \cite{framevec}
and again later in the full Standard Model context.

Having now applied the framon idea to the electroweak sector, let us push 
further and apply the same ideas to the full symmetry $u(1) \times su(2)\times 
su(3)$.  We focus first on a single framon vector.  This should be by itself 
a representation of the local symmetry $u(1) \times su(2) \times su(3)$, 
to be constructed out of the fundamental representations of the component 
symmetries, namely a phase for $u(1)$, a doublet for  $su(2)$, a triplet for 
$su(3)$.  For the product between each pair, one can take either the sum 
or the product representation.  In the electroweak theory where the pair 
$u(1) \times su(2)$ occurred, one took the product, this being the smaller 
of the two ($1 \times 2 < 1 + 2$), minimising thus the number of scalar
fields to be introduced.  The same reasoning should apply to the product
$u(1) \times su(2)$ here, and also to $u(1) \times su(3)$.  For the pair
$su(2) \times su(3)$, however, the sum representation is smaller than
the product ($2 \times 3 > 2 + 3$).  Hence, following the same line of 
reasoning, to economize on the number of scalar fields, one would opt for 
the representation ${\bf 1} \otimes ({\bf 2} \oplus {\bf 3})$. Practically, 
this means that each framon vector here is broken into 2 parts: a ``weak'' 
framon, $\phi_r, r = 1, 2$, being a doublet of $su(2)$, and a ``strong'' 
framon, $\phi_a, a = 1, 2, 3$, being a triplet of $su(3)$, with each being 
also a representation of $u(1)$, i.e.\ carrying a $u(1)$ charge.

What $u(1)$ charges should these framons carry?  By the symmetry $u(1) 
\times su(2) \times su(3)$, we mean usually the algebra which, for 
constructing the action, for example, is all that matters.  To identify 
what charges are permissible, however, we need to specify the gauge group 
\cite{ourbook}.  Again, a version of this question in the electroweak 
theory was answered above correctly by invoking minimality.  Suppose one 
takes the same attitude for the symmetry $u(1) \times su(2) \times su(3)$; 
one would then adopt as gauge group the group, denoted $U(1, 2, 3)$ say, 
which is obtained by identifying in the group $U(1) \times SU(2) \times SU(3)$ 
the following sextet of elements,
\begin{eqnarray}
\lefteqn{
(y, f, c) = (z^4 y, f, \omega c) = (z^2 y, f, \omega^2 c)}\nonumber \\
&& {} 
= (z^3 y, - f, c)= (z y, -f, \omega c) = (z^5 y, -f, \omega^2 c)\;,
\label{U23}
\end{eqnarray}
where $y$, $f$, $c$, are elements in respectively in the groups $U(1)$, 
$SU(2)$, and $SU(3)$, 
\begin{equation}
z = \exp  \pi i/3, \quad \omega=\exp 2\pi i/3.
\label{z}
\end{equation}
Note that $(\pm 1)$ and 
$(1, \omega, \omega^2)$ 
can respectively be identified with the elements of the centre of 
the groups $SU(2)$ and $SU(3)$.  Being 6-fold covered by the group 
$U(1) \times SU(2) \times SU(3)$, $U(1,2,3)$ is the ``smallest'' group with 
algebra $u(1) \times su(2) \times su(3)$ which admits as representations 
both the {\bf 2} of $su(2)$ and the {\bf 3} of $su(3)$.  Then, with 
$U(1,2,3)$ specified as the gauge group, it follows that any field in 
the theory, which has to be a representation of $U(1,2,3)$, can have only
the $u(1)$ charges $g_1 q$, with $q$ depending on the representations in 
$su(2)$ and $su(3)$ as follows \cite{ourbook}:
\begin{eqnarray}
&(1, 1); \ \ \ & q = 0 + n, \nonumber \\
&(2, 1); \ \ \ & q = \half + n, \nonumber \\ 
&(1, 3); \ \ \ & q = -\third + n, \nonumber \\
&(2, 3); \ \ \ & q = {\tfrac{1}{6}} +n,
\label{qadmit}
\end{eqnarray}
where the first number inside the brackets denotes the dimension of the
representation in $su(2)$ and the second number that in $su(3)$, and $n$
can be any integer, positive or negative.

Restricting further the representations of the framons in $U(1, 2, 3)$ to 
those with minimal $|q|$,  (i.e.\ equivalents in $U(1)$ of fundamental or 
antifundamental representations) one obtains then for the ``weak'' framon 
$q = \pm 1/2$ and for the ``strong'' framon $q =- 1/3$.  In consequence, 
they will acquire also the $\tilde{u}(1)$ charges $g_1 \tilde{q}$ with 
$\tilde{q}$ opposite in sign to $q$, i.e.\ $\tilde{q} = \mp \half$ for 
the ``weak'' framons, and $\tilde{q} = \third$ for the ``strong'' framons.

We need more than one such framon, indeed as many framons as there are frame
vectors for the symmetry $u(1) \times su(2) \times su(3)$.  We recall that,
as for the simple symmetries $su(N)$ in (\ref{Phi}) above, the ``rows'' of
the present $\Phi$ matrix, i.e.\ the $(r,a)$-th components 
(for fixed $r,a$) of 
the various frame vectors labelled by $(\tilde{r}, \tilde{a})$, 
and hence also of the framons, together should transform as 
a representation of the global symmetry $\tilde{u}(1) \times \widetilde{su}(2)
\times \widetilde{su}(3)$.  We thus have to ask again, which representation?
Minimality would suggest again $\tilde{\bf 1} \otimes (\tilde{\bf 2} 
\oplus \tilde{\bf 3})$,
but if one chooses that, the theory would just break up into 2 separate 
theories, i.e.\ the electroweak theory plus chromodynamics disjoint from one
another, which is not an interesting nor realistic situation.  
We propose therefore to 
opt instead for the all product representation $\tilde{\bf 1} \otimes 
\tilde{\bf 2} \otimes \tilde{\bf 3}$.  Although this apparently departs from 
the ``principle of minimality'' which has so far been our guideline, it does 
not affect the number of scalar ($x$-dependent) fields that have to be 
introduced which is still minimal since the symmetry under consideration 
is only global.  In any case, if the proposal of the product representation 
$\tilde{\bf 1} \otimes \tilde{\bf 2} \otimes \tilde{\bf 3}$ is accepted, 
we have finally for the full list of framon fields the following.  For the 
``weak'' framons, we have:
\begin{equation}
\phi_r^{\tilde{r} \tilde{a}} = \alpha^{\tilde{a}} \phi_r^{\tilde{r}},
    \ \ r, \tilde{r} = 1, 2, \ \  \tilde{a} = 1, 2, 3, \ \ \ \ q = 
\pm \half, \ \
    \tilde{q} = \mp \half,
\label{phirrtat}
\end{equation}
and for the ``strong'' framons,
\begin{equation}
\phi_a^{\tilde{r} \tilde{a}} = \alpha^{\tilde{r}} \phi_a^{\tilde{a}},
    \ \ \tilde{r} = 1, 2, \ \ a, 
\tilde{a} = 1, 2, 3, \ \ \ \  q = - \third, \ \
    \tilde{q} = \third.
\label{phirtaat}
\end{equation}
The quantities $\alpha^{\tilde{r}}$ and $\alpha^{\tilde{a}}$ are independent
of $x$, but $\phi_r^{\tilde{r}}$ and $\phi_a^{\tilde{a}}$ are $x$-dependent 
representing altogether $4 + 9 = 13$ complex scalar fields or 26 real fields.  
Because of the unitarity constraint (\ref{su2ortho}) for $\phi_r^{\tilde{r}}$ 
and that the determinant be real for $\phi_a^{\tilde{a}}$, not all of these 
are actually independent, and one ends up for the framons with just 21 real
field degrees of freedom.

The specific set-up (\ref{phirrtat}) and (\ref{phirtaat}) of framons is
arrived at by insisting, when faced with ambiguities, on ``minimality'',
which though appealing for economy is by no means compelling since one
cannot as yet give a physical reason why nature should opt for economy.
Indeed, we have to admit that in arriving at the above conclusion along
the lines described, we have been peeping at the physical consequences 
also, and have no doubt been influenced by physical considerations.  For
example, in deciding on which representation a framon should take in the
symmetry $u(1) \times su(2) \times su(3)$, we have tried at first the
full product ${\bf 1} \otimes {\bf 2} \otimes {\bf 3}$ but had to discard 
it for giving a physically inadmissible amount of mixing between the weak 
and strong sectors.  Also, while opting for the group $U(1,2,3)$, we knew 
already from previous analysis that it is the group that nature seems to 
prefer.  So, the above derivation through minimality is in a sense an 
afterthought.  Nevertheless, we find it interesting that by appealing to 
minimality when faced with ambiguities left open by the framon idea, one is 
led almost uniquely (i.e.\ apart from the choice of representation for the 
global symmetry already mentioned) to the only solution which seems to 
work, namely the one set out in (\ref{phirrtat}) and (\ref{phirtaat}) 
that we shall henceforth adopt.

The framon fields in (\ref{phirrtat}) and (\ref{phirtaat}) are of a form
unfamiliar at least to us, having as they do each an $x$-independent factor 
$\alpha$ which depends on symmetry indices.  We find it convenient to 
picture them as rather like the nucleon wave function in nuclear physics 
when isospin is considered as an exact global symmetry.  To describe there 
a nucleon, we have not only to give the wave function dependent on $x$ and 
on spin, but also to specify whether it is a proton or a neutron by an 
isospin factor independent both of $x$ and of spin.

For future use, we choose to normalize the $\alpha$'s, as we are free to do, 
such that
\begin{equation}
\sum_{\tilde{a}} |\alpha^{\tilde{a}}|^2 = 1, \ \ \ 
\sum_{\tilde{r}} |\alpha^{\tilde{r}}|^2 = 1.
\label{alphanorm}
\end{equation}
We shall find it convenient also to adopt on occasion a vector notation 
grouping various components of the framons, thus:
\begin{equation}
(\bphi^{\tilde{r}})_r = (\bphi_r)^{\tilde{r}} =
\phi_r^{\tilde{r}}\,, \ \ \ 
(\bphi^{\tilde{a}})_a = (\bphi_a)^{\tilde{a}} = \phi_a^{\tilde{a}}\,,
\label{vectornot}
\end{equation}
where $\bphi$ can denote a 2-dimensional vector in 
either $su(2)$ or $\widetilde{su}(2)$ space,
or a 3-dimensional vector in either $su(3)$ or $\widetilde{su}(3)$ space.
Any ambiguity which might arise in this notation can readily be resolved
by noting how the vectors are labelled and by the context.

For the weak framons $\phi_r^{\tilde{r}\tilde{a}}$, one can economize on the 
notation further by eliminating $\bphi^{\tilde{2}}$ via (\ref{su2ortho}) 
leaving only 
\begin{equation}
\alpha^{\tilde{a}} \phi_r = \alpha^{\tilde{a}} (\bphi^{\tilde{1}})_r\,, 
\label{phiw}
\end{equation} 
where $\phi_r$ can be taken with a definite charge  e.g.\
$\phi_r^{(-)}$ as we 
shall sometimes do in future.  However, to exhibit the $\widetilde{su}(2)$
invariance of the theory, it is often more convenient to keep the more
general notation with both $\tilde{r} = 1, 2$ where neither carries a 
definite $u(1)$ charge, thus:
\begin{equation}
\phi^{(\pm)}_r = \alpha_{\tilde{1}}^{(\pm)} \phi^{\tilde{1}}_r
   + \alpha_{\tilde{2}}^{(\pm)} \phi^{\tilde{2}}_r
   = \balpha^{(\pm)}\cdot \bphi_r,
\label{su2tphi}
\end{equation}
with
\begin{equation}
|\balpha^{(+)}| = |\balpha^{(-)}| = 1, \ \ \ \balpha^{(+)} \cdot
\balpha^{(-)} = 0.
\label{alpha+-}
\end{equation}

With the specification of the framon fields in addition to the gauge fields
introduced at the beginning, our list of the fundamental boson fields is 
now complete.

\setcounter{equation}{0}

\section{The Invariant Action}

Next, we turn to the construction of an invariant action for the bosonic
fields enumerated above.  Apart from Lorentz invariance, we shall require 
of course that the action be invariant under local transformations of the 
initial gauge symmetries $u(1) \times su(2) \times su(3)$.  Furthermore,
given that the global symmetries $\tilde{u}(1) \times \widetilde{su}(2)
\times \widetilde{su}(3)$ which enter in the framon fields originate only
as the choice of reference frames of which physics should presumably be
independent, we ought to require also that our action be invariant under
these as well.  We shall therefore construct an action on this basis,
which is fairly stringent, leaving rather little freedom for its choice.

As usual we write our action for the bosonic sector as a sum of 3 terms:
\begin{equation}
{\cal A}_B = {\cal A}_{GF} + {\cal A}_{KE} + \int V,
\label{calAB}
\end{equation}
with ${\cal A}_{GF}$ representing the free action for the gauge fields, 
and ${\cal A}_{KE}$ and $V$ respectively the kinetic energy and the 
potential of self-interaction for the scalar framons.

Explicitly, for ${\cal A}_{GF}$, we write as usual: 
\begin{equation}
{\cal A}_{GF} = - \frac{1}{4} \int\! d^4 x \,F_{\mu\nu} F^{\mu\nu}
                - \frac{1}{4} \int\! d^4 x\, \Tr(G_{\mu\nu} G^{\mu\nu})
                - \frac{1}{4} \int\! d^4 x\, \Tr(H_{\mu\nu} H^{\mu\nu})\,,
\label{calAGF}
\end{equation}
with
\begin{eqnarray}
F_{\mu\nu} & = & \partial_\nu A_\mu - \partial_\mu A_\nu\,,  \nonumber \\
G_{\mu\nu} & = & \partial_\nu B_\mu - \partial_\mu B_\nu + 
ig_2\,[B_\mu,B_\nu]\,,   \nonumber \\
H_{\mu\nu} & = & \partial_\nu C_\mu - \partial_\mu C_\nu + 
ig_3\,[C_\mu,C_\nu]\,,   \nonumber \\
\label{gaugefields}
\end{eqnarray}
for respectively the $u(1), su(2)$ and $su(3)$ components. This action 
${\cal A}_{GF}$ is of course constructed to be invariant under the local 
symmetries.  It is invariant also under the global symmetries  $\tilde{u}(1) 
\times \widetilde{su}(2) \times \widetilde{su}(3)$, trivially since on 
these it does not depend.

Next, for the kinetic energy term ${\cal A}_{KE}$ we write as usual:
\begin{equation}
{\cal A}_{KEW} = \int \sum_{r,\tilde{r},\tilde{a}} \left[ \left(\partial_\mu
    - i g_1q  A_\mu - i g_2 B_\mu \right)_{rs}
      \phi_s^{\tilde{r} \tilde{a}} \right]^* \times \left[ \rm{h.c.} \right],
\label{calAKEW}
\end{equation}
for the weak framons $\phi_r^{\tilde{r} \tilde{a}}$, where $q$ denotes the 
$u(1)$ charge operator, and
\begin{equation}
{\cal A}_{KES} = \int \sum_{a,\tilde{r},\tilde{a}} \left[ \left(\partial_\mu
  +i \frac{g_1}{3}A_\mu  - i g_3 C_\mu \right)_{ab}
     \phi_b^{\tilde{r} \tilde{a}} \right]^* \times \left[ \rm{h.c.} \right],
\label{calAKES}
\end{equation}
for the strong framons $\phi_a^{\tilde{r} \tilde{a}}$.  These kinetic energy 
terms are by construction invariant under both the local and the global 
symmetries.

By virtue of the conditions (\ref{phirrtat}), the weak framon term reduces to:
\begin{equation}
{\cal A}_{KEW} = \int \sum_{\tilde{r}} (D_\mu \bphi^{\tilde{r}})^\dagger
    (D^\mu \bphi^{\tilde{r}})\,,
\label{calAKEW1}
\end{equation}
with
\begin{equation}
D_\mu \bphi^{\tilde{r}} = \left( \partial_\mu - i g_1 q
   A_\mu - i g_2 B_\mu \right) \bphi^{\tilde{r}}
\label{coderivW}
\end{equation}
while by (\ref{phirtaat}) the strong framon term reduces to:
\begin{equation}
{\cal A}_{KES} = \int \sum_{\tilde{a}} (D_\mu \bphi^{\tilde{a}})^\dagger
    (D^\mu \bphi^{\tilde{a}})\,,
\label{calAKES1}
\end{equation}
with
\begin{equation}
D_\mu \bphi^{\tilde{a}} = \left( \partial_\mu +i \frac{g_1}{3} A_\mu
     - i g_3 C_\mu \right) \bphi^{\tilde{a}}.
\label{coderivS}
\end{equation}
Further, by virtue of the conditions (\ref{su2ortho}) and (\ref{phiw}), 
the kinetic energy term for weak framons reduces to just:
\begin{equation}
{\cal A}_{KEW} = 2 \int (D_\mu \bphi)^\dagger (D^\mu \bphi)\,,
\label{calAKEWSM}
\end{equation}
which, if we choose $\bphi = \bphi^{(-)}$, is the same as in standard 
electroweak theory, apart from an unimportant factor 2.

Lastly, to construct the general interaction potential for framons with 
the required invariance under both the local and global symmetries, we 
take a sum of all terms up to fourth order (for renormalizability) in the 
framon fields, contracting the indices in all possible ways.  We obtain 
the following.:
\begin{equation}
V[\Phi] = V_W[\Phi] + V_S[\Phi] + V_{WS}[\Phi]\,,
\label{VPhi}
\end{equation}
where $V_W$ involves only the weak framons, $V_S$ only the strong framons,
and $V_{WS}$ both.  Explicitly, for the weak sector, we have:
\begin{eqnarray}
V_W[\Phi] & = & - \mu'_W \sum_{r,\tilde{r},\tilde{a}}
      \phi_r^{\tilde{r} \tilde{a} *} \phi_r^{\tilde{r} \tilde{a}}
    + \lambda'_W \left[ \sum_{r,\tilde{r},\tilde{a}}
      \phi_r^{\tilde{r} \tilde{a} *} \phi_r^{\tilde{r} \tilde{a}} \right]^2
    +  \kappa_{1W} \sum_{r,s,\tilde{r},\tilde{s},\tilde{a},\tilde{b}}
      \phi_r^{\tilde{r} \tilde{a} *} \phi_r^{\tilde{r} \tilde{b}}
      \phi_s^{\tilde{s} \tilde{b} *} \phi_s^{\tilde{s} \tilde{a}}
      \nonumber \\
  & + & \kappa_{2W} \sum_{r,s,\tilde{r},\tilde{s},\tilde{a},\tilde{b}}
      \phi_r^{\tilde{r} \tilde{a} *} \phi_r^{\tilde{s} \tilde{a}}
      \phi_s^{\tilde{s} \tilde{b} *} \phi_s^{\tilde{r} \tilde{b}}
    + \kappa_{3W} \sum_{r,s,\tilde{r},\tilde{s},\tilde{a},\tilde{b}}
      \phi_r^{\tilde{r} \tilde{a} *} \phi_s^{\tilde{r} \tilde{a}}
      \phi_s^{\tilde{s} \tilde{b} *} \phi_r^{\tilde{s} \tilde{b}},
\label{VPhiW}
\end{eqnarray}
for the strong sector, we have:
\begin{eqnarray}
V_S[\Phi] & = & - \mu'_S \sum_{a,\tilde{r},\tilde{a}}
     \phi_a^{\tilde{r} \tilde{a} *} \phi_a^{\tilde{r} \tilde{a}}
    + \lambda'_S \left[ \sum_{a,\tilde{r},\tilde{a}}
      \phi_a^{\tilde{r} \tilde{a} *} \phi_a^{\tilde{r} \tilde{a}} \right]^2
    + \kappa_{1S} \sum_{a,b,\tilde{r},\tilde{s},\tilde{a},\tilde{b}}
      \phi_a^{\tilde{r} \tilde{a} *} \phi_a^{\tilde{s} \tilde{a}}
      \phi_b^{\tilde{s} \tilde{b} *} \phi_b^{\tilde{r} \tilde{b}}
       \nonumber \\
  & + & \kappa_{2S} \sum_{a,b,\tilde{r},\tilde{s},\tilde{a},\tilde{b}}
      \phi_a^{\tilde{r} \tilde{a} *} \phi_a^{\tilde{r} \tilde{b}}
      \phi_b^{\tilde{s} \tilde{b} *} \phi_b^{\tilde{s} \tilde{a}}
    + \kappa_{3S} \sum_{a,b,\tilde{r},\tilde{s},\tilde{a},\tilde{b}}
      \phi_a^{\tilde{r} \tilde{a} *} \phi_b^{\tilde{r} \tilde{a}}
      \phi_b^{\tilde{s} \tilde{b} *} \phi_a^{\tilde{s} \tilde{b}},
\label{VPhiS}
\end{eqnarray}
and for the interaction between the two, we have:
\begin{eqnarray}
\hspace*{-10mm}V_{WS}[\Phi] & = &
     \!\! \nu_{11} \sum_{r,a,\tilde{r},\tilde{s},\tilde{a},\tilde{b}}
      \phi_r^{\tilde{r} \tilde{a} *} \phi_r^{\tilde{r} \tilde{a}}
      \phi_a^{\tilde{s} \tilde{b} *} \phi_a^{\tilde{s} \tilde{b}}
    + \nu_{21} \sum_{r,a,\tilde{r},\tilde{s},\tilde{a},\tilde{b}}
      \phi_r^{\tilde{r} \tilde{a} *} \phi_r^{\tilde{r} \tilde{b}}
      \phi_a^{\tilde{s} \tilde{b} *} \phi_a^{\tilde{s} \tilde{a}}
      \nonumber \\
  & + & \!\!\nu_{12} \sum_{r,a,\tilde{r},\tilde{s},\tilde{a},\tilde{b}}
      \phi_r^{\tilde{r} \tilde{a} *} \phi_r^{\tilde{s} \tilde{a}}
      \phi_a^{\tilde{s} \tilde{b} *} \phi_a^{\tilde{r} \tilde{b}}
    + \nu_{22} \sum_{r,a,\tilde{r},\tilde{s},\tilde{a},\tilde{b}}
      \phi_r^{\tilde{r} \tilde{a} *} \phi_r^{\tilde{s} \tilde{b}}
      \phi_a^{\tilde{s} \tilde{b} *} \phi_a^{\tilde{r} \tilde{a}}\,.
\label{VPhiWS}
\end{eqnarray}

Imposing now the conditions (\ref{su2ortho}) and (\ref{phiw}) eliminating
thus $\phi_r^{\tilde{2}}$ in terms of $\phi_r^{\tilde{1}} = \phi_r$, we
find that the $\kappa$ terms in $V_W$ are of the same form as the $\lambda$
term so that the whole of $V_W$ reduces to:
\begin{equation}
V_W[\Phi] = - \mu_W |\bphi|^2 + \lambda_W (|\bphi|^2)^2,
\label{VPhiW2}
\end{equation}
i.e.\ the same as in the standard electroweak theory, only with $\mu_W =
2 \mu'_W$, and $\lambda_W = 4 \lambda'_W + 4 \kappa_{1W} + 2 \kappa_{2W}
+ 2 \kappa_{3W}$.  Imposing next the conditions (\ref{phirtaat}) on $V_S$, we
find that the $\kappa_{1S}$ term becomes the same as the $\lambda_S$ term
while the $\kappa_{2S}$ and $\kappa_{3S}$ terms take on the same form, so
that the whole of $V_S$ reduces to:
\begin{equation}
V_S[\Phi] = - \mu_S \sum_{a,\tilde{a}} (\phi_a^{\tilde{a}*}\phi_a^{\tilde{a}})
    + \lambda_S \left[ \sum_{a, \tilde{a}} (\phi_a^{\tilde{a}*}
    \phi_a^{\tilde{a}}) \right]^2 + \kappa_S \sum_{a,b,\tilde{a},\tilde{b}}
    (\phi_a^{\tilde{a}*} \phi_a^{\tilde{b}})
    (\phi_b^{\tilde{b}*} \phi_b^{\tilde{a}}),
\label{VPhiS2}
\end{equation}
which is the same as the framon potential for the pure $su(3)$ theory given
in \cite{framevec}, only with $\mu_S = \mu'_S$, $\lambda_S = \lambda'_S 
+ \kappa_{1S}$, and $\kappa_S = \kappa_{2S} + \kappa_{3S}$.  Similarly, 
imposing both (\ref{phirrtat}) and (\ref{phirtaat}), one obtains that in 
$V_{WS}$, the $\nu$ terms are equal in form in pairs, giving:
\begin{equation}
V_{WS}[\Phi] = \nu_1 |\bphi|^2 \sum_{a,\tilde{a}} \phi_a^{\tilde{a} *}
    \phi_a^{\tilde{a}} + \nu_2 |\bphi|^2 \sum_a \left| \sum_{\tilde{a}}
    (\alpha^{\tilde{a} *} \phi_a^{\tilde{a}})\, \right|^2,
\label{VPhiWS2}
\end{equation}
with $\nu_1 = 2 \nu_{11} + \nu_{12}$ and $\nu_2 = 2 \nu_{21} + \nu_{22}$.

As far as we can see, given the criteria for invariance under the prescribed
symmetries and renormalizability, the potential $V[\Phi]$ so constructed 
is unique.

\setcounter{equation}{0}

\section{The Framon Potential Vacuum}

Our next job is to find the minima of the framon potential to identify
the vacuum which, as we shall see, has some intriguing properties with
interesting consequences.  The problem is of course in principle entirely
soluble just by differentiating the potential with respect to the various
components of the framon fields, but the potential being rather complicated,
it pays first to examine the problem qualitatively to see what answer one 
might expect.  To do so, we find it convenient to express the potential in 
terms of the 3-vectors $\balpha$ and $\bphi_a, a = 1, 2, 3$, thus
\begin{eqnarray}
V[\Phi] & = & - \mu_W |\bphi|^2 + \lambda_W (|\bphi|^2)^2 \nonumber \\
         &   & - \mu_S \sum_a |\bphi_a|^2 + \lambda_S \left( \sum_a
         |\bphi_a|^2 \right)^2 + \kappa_S \sum_{a,b} |(\bphi_a^* \cdot
         \bphi_b)|^2
         \nonumber \\
         &   & + \nu_1 |\bphi|^2 \sum_a |\bphi_a|^2
               - \nu_2 |\bphi|^2 \sum_a |(\balpha^* \cdot \bphi_a)|^2.
\label{VPhivec}
\end{eqnarray}
Notice that $\balpha$ and $\bphi_a, a = 1, 2, 3$ here are vectors in 
$\widetilde{su}(3)$ space, not in $su(3)$ space as were the vectors
$\bphi^{\tilde{a}}$ in the expression (\ref{calAKES1}) for the kinetic 
energy term ${\cal A}_{KE}$.  The signs of the coefficients $\mu, \lambda,
\kappa$ and $\nu$ are all in principle arbitrary as far as invariance is
concerned, but we shall choose them all to be positive for the present 
discussion.  The chosen signs for the $\mu$'s and $\lambda$'s are the 
same as for the standard electroweak theory and are not therefore new. 
The signs chosen for $\nu_1$ and $\nu_2$ are not really necessary for 
our purpose as far as we can see, but are chosen such only for ease of 
presentation.  This leaves then the sign chosen for $\kappa_S$ as 
the only genuinely new assumption.

To study the minimization problem of the above potential (\ref{VPhivec}),
we shall proceed in 2 steps.  We shall consider first the 2 parts of the 
potential labelled previously as $V_W$ and $V_S$, the former depending 
only on the weak framon $\bphi$ and the latter depending only on the 
strong framons $\bphi_a$.  Then secondly, we shall examine the effect of 
the remaining terms linking the weak and strong sectors, namely $\nu_1$ 
and $\nu_2$, which we shall treat as disturbances on the purely weak and 
purely strong potentials.

The weak potential comprising the first 2 terms in (\ref{VPhivec}) is the
same as in standard electroweak theory with minimum at $|\bphi|^2  = \mu_W/
(2 \lambda_W)$ of which nothing more needs to be said.  The strong potential
consisting of the following 3 terms in (\ref{VPhivec}) has more intricate
features.  We notice first that the $\kappa_S$ term is the only term which 
depends on the relative orientations among the vectors $\bphi_a, a = 1, 2, 3$, 
the other 2 terms $\mu_S$ and $\lambda_S$ being functions only of their lengths
$|\bphi_a|$.  Hence one can minimize first the $\kappa_S$ term with respect
to the orientations of the vectors, and obtain for $\kappa_S$ positive that
the 3 vectors $\bphi_a, a = 1, 2, 3$ at minimum are mutually orthogonal.
This means that at minimum one can omit from the $\kappa_S$ term all those
terms with $a \neq b$ leaving:
\begin{equation}
V_{SR}[\Phi] = - \mu_S \sum_a |\bphi_a|^2 + \lambda_S \left( \sum_a
         |\bphi_a|^2 \right)^2 + \kappa_S \sum_a (|\bphi_a|^2)^2,
\label{VPhiSR}
\end{equation}
depending only on the lengths $|\bphi_a|$, and minimising it with respect 
to these easily gives:
\begin{equation}
|\bphi_1|^2 = |\bphi_2|^2 = |\bphi_3|^2 = \frac{ \mu_S}{6 \lambda_S 
+ 2\kappa_S}\;. 
\label{bphiamin}
\end{equation}
In other words, at minimum, the framon fields $\bphi_a$ when normalized 
would form an orthonormal triad, exactly as what one would expect for 
frame vectors.  A little more detail for the properties of the strong 
potential $V_S$ can be found in \cite{framevec}.

We turn now to the terms $\nu_1$ and $\nu_2$ linking the weak and strong
sectors.  The $\nu_1$ term affects only the squared length $|\bphi|^2$ of 
the weak framon and the squared length: 
\begin{equation}
\zeta_S^2  = \sum_a |\bphi_a|^2
\label{zeta}
\end{equation}
of the strong, and present thus no very new features.  We shall thus first 
focus on the influence of the $\nu_2$ term on the framons in the strong 
sector.  We shall regard this effect as perturbative, seeing that the $\nu$ 
terms as compared with the terms in the strong potential are of the order 
$|\bphi|^2/|\bphi_a|^2$, which we have reason to believe is small.  However, 
this is just a matter of convenience, for most of what we conclude would 
still hold even if this turns out not to be true.

We recall that, because of the $\kappa_S$ term in the strong potential, 
its minimization implies that the framons $\bphi_a$ form an orthogonal 
triad all of the same length. We ask now what will happen when we 
minimize the $\kappa_S$ and  $\nu_2$ terms together.  Consider first the 
situation when these framons are kept still having the same length, thus
allowing only their orientations to vary.  For the $+$ sign of the $\nu_2$
term that we have chosen, this term is smallest when the framons $\bphi_a$
are all aligned with the vector $\balpha$, but this is opposed by the
$\kappa_S$ term which, to attain its smallest value, would want instead
the framons to be mutually orthogonal.  The result of minimising the 2
terms together would thus be a compromise where the originally orthogonal
triad is squeezed together a little towards the vector $\balpha$ which,
by the symmetry of the problem, would be placed symmetrically with respect
to the triad.  The amount of distortion $\delta$ to the orthogonal triad 
would be of the order $|\bphi|^2/|\bphi_a|^2$.

Next, consider the opposite situation when the framons are kept mutually 
orthogonal but allowed to change their lengths relative to one another. 
We recall first that the $\mu_S$ and $\lambda_S$ terms of the purely 
strong potential depend only on $\zeta_S$, not on how it is distributed 
among the three lengths $|\bphi_a|$.  It was the $\kappa_S$ term whose 
minimization gave the result that the 3 lengths should be equal.  However, 
this is again opposed by the $\nu_2$ linkage term which, to achieve its 
smallest value, would prefer to have all the weight of $\zeta_S$ attributed 
to just one of the framons $\bphi_a$ and $\balpha$ to be aligned with it. 
Minimizing the 2 terms together would thus once more lead to a compromise 
where the minimum would be displaced from the symmetrical point favoured by 
the purely strong potential to a point with the lengths of the $\bphi_a$'s 
differing from one another by amounts of the order $\zeta_S \delta$, and 
with the vector $\balpha$ aligned with the longest.

From the conclusion in the above 2 extreme situations, it is then not
difficult to visualize what will happen in the actual situation when the
framons are allowed both to deviate from mutual orthogonality and to 
change in length relative to one another.  The original orthogonal triad
will be squeezed slightly by an amount of order $|\bphi|^2/|\bphi_a|^2$, 
while at the same time the framons will deviate in relative length from one 
another by amounts of the same order, with the vector $\balpha$ snuggling 
up to the longest among them.  Since the squeezing of the orthogonal triad
has a similar effect to giving the framons different lengths in reducing 
the value of the potential, one can imagine a kind of trade-off between 
the two, or that the minimum of the potential is degenerate with respect 
to the amount of squeeze on the triad so long as the effect is compensated 
by a simultaneous change in the relative framon lengths in a prescribed 
manner.  The degeneracy would be 2-dimensional, there being 2 angles in the 
triad to be squeezed or 2 relative lengths of vectors to be varied.

Next, we ask how the vector $\balpha$ will change when the relative lengths 
of the strong framons vary.  We note first that when the strong framons 
$\bphi_a$ are strictly orthogonal to one another and have the same length, 
then the $\nu_2$ term will have no preference for the direction of $\balpha$, 
the term being proportional to the sum of the squares of the direction 
cosines of $\balpha$ and therefore independent of the direction 
of $\balpha$.  However, as soon as the framons are allowed to have different 
lengths, then $\balpha$ will flop to align itself with the longest one for 
the $\nu_2$ term will then acquire its minimum value.  But if we now allow 
the orthogonality also to be distorted, then $\balpha$ can no longer align 
itself exactly with the longest framon for its inner products with the 
shorter framons will now also contribute.  Hence, $\balpha$ will end up 
quite closely but not exactly aligned with the longest framon.  Nevertheless, 
for small squeezing angles, the system is close to being unstable so that 
even a small change in relative lengths of the framons from identity would 
be enough to bring about a large change in the direction of $\balpha$.  In 
other words, if we choose to write:
\begin{equation}
(|\bphi_1|, |\bphi_2|, |\bphi_3|) = \zeta_S (x, y, z); \ \ \
    x^2 + y^2 + z^2 = 1,
\label{intxyz}
\end{equation}
considering ${\bf r} = (x, y, z)$ itself as a unit vector so that changes 
in the relative lengths of $\bphi_a$'s can be pictured as a rotation of 
this vector, then our previous conclusion can be restated as saying that 
a small rotation of ${\bf r}$ will lead to a large rotation of $\balpha$,
or that the directions of ${\bf r}$ and $\balpha$ are coupled such that
rotations in ${\bf r}$ will become magnified in $\balpha$.  The smaller 
the value of the squeezing angle $\delta$, the closer is the system to 
instability and the more effective is the magnification.

All these properties of the vacuum can, of course, be derived just by 
minimising the framon potential (\ref{VPhi}).  However, the algebraic 
complications, due mainly to the deviations of the vacuum from
orthonormality,
tend to obscure the basic structure of the calculation.  For 
clarity of presentation, therefore, we shall work out in full here only 
a simplified version of it where there are only 2 colours (and hence only 
2 fermion generations as the dual of colour) instead of the actual 3. 
This will contain in paraphrase already all the essential features in the 
actual 3-generation problem. 
The calculation on the vacuum and its subsequent applications for the
actual 3-generation case will, in this paper, be carried only to such
an approximation and extent as to satisfy ourselves that it holds no
surprises beyond what is discovered from the 2-generation model.  A
full calculation is under way and will be reported on separately
if found necessary.

By the 2-G (2 colours and 2 generations) model of the strong sector, we 
mean a theory with a local $su(2)$ symmetry representing colour plus a global 
$\widetilde{su}(2)$ symmetry representing generations.  Such a theory can, 
as described at the beginning of Section 2, be formulated via (\ref{su2ortho}) 
with a smaller number of framons, as with the electroweak theory.  However, 
in order to simulate better the actual 3-G case, we shall take the framon 
fields here as the components of a $2 \times 2$ matrix $\Phi$, satisfying 
only the constraint that the determinant be real.

Our convention, we recall is that the columns of $\Phi$ transform as doublets 
under local $su(2)$ while its rows transform as anti-doublets under global 
$\widetilde{su}(2)$.  We can then choose to work in the (local) gauge where 
the first column of $\Phi$ points in the first direction and is real, the 3 
degrees of freedom in $su(2)$ being just sufficient for us to do so. The 
condition that the determinant of $\Phi$ be real then implies the second 
diagonal element to be real also.  We shall refer henceforth to this as the 
triangular gauge, in which $\Phi$ can be parametrized as:
\begin{equation}
\Phi = \left( \begin{array}{cc}
    X \cos \delta & X \sin \delta\; e^{i \phi} \\  0 & Y
    \end{array} \right).
\label{2Gtriang}
\end{equation}
This $\Phi$ representing the strong framon fields is coupled in the potential
(\ref{VPhi}) to the weak sector via the vector $\balpha$ which we parametrize 
in this gauge as:
\begin{equation}
\balpha = \left( \begin{array}{ll}
    \cos \alpha\; e^{-i \beta} \\ \sin \alpha\; e^{-i \gamma}
    \end{array} \right).
\label{2Galpha}
\end{equation}
We can then write the potential as:
\begin{eqnarray}
V[\Phi] & = & - \mu_W \zeta_W^2 + \lambda_W \zeta_W^4 \nonumber \\
    & & - \mu_S (X^2 + Y^2) + \lambda_S (X^2 + Y^2)^2
        + \kappa_S (X^4 + Y^4 + 2 X^2 Y^2 \sin^2 \delta) \nonumber \\
    & & + \nu_1 \zeta_W^2 (X^2 + Y^2) \nonumber \\
    & & - \nu_2 \zeta_W^2 [(X \cos \delta \cos 
\alpha + X \sin \delta
          \sin \alpha \cos \theta)^2\!+\!X^2 \sin^2 \delta \sin^2 \alpha
          \sin^2 \theta] \nonumber \\
    & & - \nu_2 \zeta_W^2 Y^2 \sin^2 \alpha,
\label{2GVreduced}
\end{eqnarray}
as a function of the 6 variables $X, Y, \delta, \alpha,$ 
\begin{equation}
\zeta_W^2 =|\bphi|^2, \quad {\rm and} \quad
\theta = \phi + \beta - \gamma.
\label{theta}
\end{equation}

The vacuum of the potential is to be obtained by putting the derivatives
of $V$ with respect to all 6 variables to zero.  First, differentiating
with respect to $\theta$, we easily obtain that the minimum occurs at
$\theta = 0$.  Differentiating with respect to the remaining 5 variables,
we then obtain, for $\theta = 0$:
\begin{eqnarray}
\zeta_S^2 & = & \frac{2 \mu_S + \nu_2 \zeta_W^2 [\cos^2 (\alpha - \delta)
    + \sin^2 \alpha] - 2 \nu_1 \zeta_W^2}{4 \lambda_S + 2 \kappa_S
    (1 + \sin^2 \delta)}, \\
\label{V2GdzetaS}
\Delta & = & \left( \frac{\nu_2 \zeta_W^2}{2 \kappa_S \zeta^2} \right)
    \frac{\cos (2 \alpha - \delta)}{\cos \delta}, \\ 
\label{V2GdDelta}
\sin 2 \delta & = & 2 \left( \frac{\nu_2 \zeta_W^2}{2 \kappa_S \zeta^2} \right)
    \frac{\sin 2 \alpha}{1 + \Delta}, \\
\label{V2Gddelta}
\sin 2 (\alpha - \delta) & = & \left( \frac{1 - \Delta}{1 + \Delta} \right)
    \sin 2 \alpha, 
\label{V2Gdalpha}
\end{eqnarray}
and
\begin{equation}
\zeta_W^2 = \frac{2 \mu_W + \nu_2 \zeta_S^2 [( 1 + \Delta) \cos^2 (\alpha
    - \delta)  + (1 - \Delta) \sin^2 \alpha] - 2 \nu_1 \zeta_S^2}{4 \lambda_W},
\label{V2GdzataW}
\end{equation}
where we have written:
\begin{equation}
X=\zeta_S x, \ \ \ Y=\zeta_S y,
\label{xy2G}
\end{equation}
with
\begin{equation}
\Delta = x^2 - y^2, \ \ \ x^2 +y^2 =1.
\label{Delta}
\end{equation}

Interestingly, these equations can be solved explicitly as follows.  We 
notice first that the equation (\ref{V2Gdalpha}) is dependent on the 2
equations before it, and can thus be omitted.  This leaves then only 4 
equations in all for our 5 unknowns $\zeta_S, \zeta_W, \Delta, \delta, 
\alpha$, meaning thus that the vacuum has a 1-dimensional degeneracy. 
Next, we see that the 2 equations (\ref{V2GdDelta}) and 
(\ref{V2Gddelta}) are solved by:
\begin{equation}
\sin \delta = \sqrt{\frac{R^2 - \Delta^2}{1 - \Delta^2}}\,, \ \ \
    \sin \alpha = \sqrt{\frac{(R - \Delta)(1 + R)}{2 R (1 - \Delta)}}\,,
\label{2Gsolvac}
\end{equation}
in terms of $\Delta$, or alternatively by:
\begin{equation}
\Delta = \frac{R(R + \cos 2 \alpha)}{1 + R \cos 2 \alpha} \ \ \ 
   \sin \delta = \frac{R \sin 2 \alpha}{\sqrt{1 + R^2 + 2R \cos 2 \alpha}},
\label{2Gsolvaca}
\end{equation}
in terms of $\alpha$, with:
\begin{equation}
R = \frac{\nu_2 \zeta_W^2}{2 \kappa_S \zeta_S^2}\,.
\label{Rdef}
\end{equation}
Substitution of these results into the remaining equations then gives 
after some algebra:
\begin{eqnarray}
\zeta_S^2 & = & \frac{\mu_S}{2 \lambda_S + 2 R \kappa_S (\nu_1/\nu_2)
    + \kappa_S(1 - R)} \label{2GsolvzetaS} \\
\zeta_W^2 & = & \frac{ 2 \mu_W + \nu_2 \zeta_S^2 (1 + R) - 2 \nu_1 \zeta_S^2}
    {4 \lambda_W}  \label{2GsolvzetaW} \\
R & = & \frac{ 2 \mu_W \lambda_S \nu_2 + \kappa_S \nu_2 \mu_W + \nu_2^2
    - 2 \nu_1 \nu_2}{8 \lambda_W \mu_S \kappa_S - 2 \kappa_S \mu_W \nu_1
    + \kappa_S \nu_2 \mu_W - \nu_2^2}\;,
\label{2GsolvR}
\end{eqnarray}
where we notice that $R$ is independent of $\Delta$ or $\alpha$, i.e.\ 
constant 
over the degenerate vacuum, and hence so are also $\zeta_S$ and $\zeta_W$ by 
(\ref{2GsolvzetaS}) and (\ref{2GsolvzetaW}).  These results confirm the
conclusions of the qualitative arguments above that both the deviations
$\Delta$ and $\delta$ of the framons from orthonormality are proportional 
to the supposedly small parameter $R$, and that changes in $\Delta$ and/or 
$\delta$ of order $R$ are enough to produce changes in $\alpha$ of order 
unity.  The solution (\ref{2Gsolvaca}) in terms of $\alpha$ is particularly
convenient in that it can be extended to all $\alpha$ ranging from 0 to 
$2 \pi$, i.e.\ over all 4 quadrants.  For $\alpha = n \pi/ 2$, i.e.\ when
the vector $\balpha$ is pointing along any of the coordinate axes, then
$\delta = 0$ or that the framons are orthogonal, but are not of the same
length, $\Delta = \pm R$.  In between, the framons deviate from 
orthogonality with $\delta > 0$ for $\balpha$ pointing in the first and 
third quadrants, but $\delta < 0$ in the second and the fourth. 

The analogous problem in the actual 3-G case, though much more complicated,
can be worked out along similar lines.  We choose again to work in the
triangular gauge where the framon field can be parametrized as:
\begin{equation}
\Phi = \left( \begin{array}{ccc}
    X \cos \delta_1 & X \sin \delta_1 \sin \gamma\; e^{i \chi_3} &
      X \sin \delta_1 \cos \gamma\; e^{i \chi_2} \\
    0 & Y \cos \delta_2 & Y \sin \delta_2\; e^{i \chi_1} \\
    0 & 0 & Z \end{array} \right)\,,
\label{3Gtriang}
\end{equation}
and the vector $\balpha$ as:
\begin{equation}
\balpha = \left( \begin{array}{l} \cos \theta\; e^{-i \beta_1} \\
    \sin \theta \sin \phi\; e^{-i \beta_2} \\
    \sin \theta \cos \phi\; e^{-i \beta_3}
    \end{array} \right).
\label{3Galpha}
\end{equation}
Substituting these into the formula (\ref{VPhi}) one obtains for the framon
potential an expression:
\begin{eqnarray}
V[\Phi] &\!\!\!=\!\!\!& -\mu_W \zeta_W^2 + \lambda_W \zeta_W^4 \nonumber \\
&& -\mu_S (X^2 + Y^2 + Z^2) + \lambda_S (X^2 + Y^2 + Z^2)^2 \nonumber
\\
&& + \kappa_S \{ X^4 + Y^4 + Z^4 + 2 X^2Y^2 \sin^2 \delta_1 \sin^2
\gamma \cos^2 \delta_2 \nonumber \\
&&+ 2 X^2 Y^2 \sin^2 \delta_1 \cos ^2 \gamma
\sin^2 \delta_2 
+ 2 X^2 Z^2 \sin^2 \delta_1 \cos^2 \gamma \nonumber \\
&& + 2 Y^2 Z^2 \sin^2 \delta_2 
+ X^2 Y^2 \sin^2 \delta_1 \sin 2 \gamma \sin
2 \delta_2 \cos (\theta_1\!+\!\theta_2\!-\!\theta_3)\} \nonumber \\
&& +\nu_1 \zeta_W^2 (X^2 + Y^2 + Z^2) \nonumber \\
&&-\nu_2 \zeta_W^2 \{X^2 (\cos^2 \delta_1 \cos^2 \theta + \sin^2
\delta_1 \sin^2 \gamma \sin^2 \theta \sin^2 \phi \nonumber \\
&&+\sin^2 \delta_1
\cos^2 \gamma \sin^2 \theta \cos^2 \phi 
+ \half \sin 2 \delta_1 \sin \gamma \sin 2 \theta \sin \phi \cos
\theta_2 \nonumber \\
&&+ \half \sin 2 \delta_1 \cos \gamma \sin 2 \theta \cos \phi
\cos \theta_3 
+ \half \sin^2 \delta_1 \sin 2 \gamma \sin^2 \theta \sin 2
\phi \cos (\theta_2\!-\!\theta_3)) \nonumber \\
&& + Y^2 ( \cos^2 \delta_2 \sin^2 \theta \sin^2 \phi\!+\!\sin^2 \delta_2
\sin^2 \theta \cos^2 \phi\!+\!\half \sin 2 \delta_2 \sin^2 \theta \sin 2
\phi \cos \theta_1) \nonumber \\
&& + Z^2 \sin^2 \theta \cos^2 \phi \}
\label{3GVreduced}
\end{eqnarray}
in terms of the 12 variables $X, Y, Z,\delta_1,  \delta_2, \gamma, \theta, 
\phi$, $\theta_1, \theta_2, \theta_3$ and $\zeta_W$ with:
\begin{equation}
\theta_1 = - \chi_1 - \beta_2 + \beta_3, \ \ \ 
\theta_2 = - \chi_3 - \beta_1 + \beta_2, \ \ \ 
\theta_3 = - \chi_2 - \beta_1 + \beta_3.
\label{thetai}
\end{equation}

Again, the vacuum is to be given by putting to zero the derivatives of the 
framon potential with respect to all these 12 variables.  The resulting 
equations are quite complicated and have been worked out so far
only to first order 
in $\zeta_W^2/\zeta_S^2$.  First, one finds from the derivatives with 
respect to $\theta_i$ that these variables should vanish at the minimum, 
thus giving:
\begin{equation}
\chi_1 = \beta_3 - \beta_2, \ \ \  \chi_2 = \beta_3 - \beta_1, \ \ \
    \chi_3 = \beta_2 - \beta_1,
\label{chii}
\end{equation}
and eliminating the $\theta_i$'s from the system.  The remaining equations
give first $\phi = \gamma$, which we use to eliminate $\phi$, leaving the 
following 8 equations.
\begin{eqnarray}
\zeta_S^2 & = &\frac{3 \mu_S + \nu_2 \zeta_W^2 -3 \nu_1 \zeta_W^2}
    {6 \lambda_S + 2 \kappa_S}\,, 
\label{3GdzetaS}\\
\Delta_1 & = & \frac{\nu_2 \zeta_W^2}{2 \kappa_S \zeta_S^2}\,
    (\cos^2 \theta - \sin^2 \theta\, \sin^2 \gamma)\,, 
\label{3GdDelta1} \\
\Delta_2 & = & - \frac{\nu_2 \zeta_W^2}{2 \kappa_S \zeta_S^2}
    \sin^2 \theta\, \cos 2 \gamma\,, 
\label{3GdDelta2} \\
\delta_1 & = & \frac{3 \nu_2 \zeta_W^2}{4 \kappa_S \zeta_S^2} \sin 2
   \theta\,, 
\label{3Gddelta1} \\
\delta_2 & = & \frac{3 \nu_2 \zeta_W^2}{4 \kappa_S \zeta_S^2}
    \sin^2 \theta\, \sin 2 \gamma\,, 
\label{3Gddelta2} \\
\delta_1 & = & \frac{3}{2} \tan 2 \theta (\Delta_1 + \Delta_2 \cos^2 \gamma
    - \frac{1}{3} \delta_2 \sin 2 \gamma)\,, 
\label{3Gddelta1a} \\
\delta_2 & = & - \frac{3}{2} \Delta_2 \tan 2 \gamma\,, 
\label{3Gddelta2a} \\
\zeta_W^2 & =& \frac{1}{6 \lambda_W} \{
3\mu_W-3\nu_1 \zeta_S^2 +\nu_2 \zeta_S^2 
(1 +2 \Delta_1 -3\Delta_1 \sin^2 \theta +\Delta_2  \nonumber \\
&& - 3\Delta_2 \sin^2 \theta \cos^2 \gamma +\delta_1 \sin 2
\theta + \delta_2 \sin 2 \gamma \sin^2 \theta) \}\,, 
\label{3GdzetaW}
\end{eqnarray}
where we have written, in analogy to the 2-G model:
\begin{equation}
X = \zeta_S x, \ \ \ Y = \zeta_S y, \ \ \ Z = \zeta_S z,
\label{xyzdef}
\end{equation}
with
\begin{equation}
\Delta_1 = x^2 - y^2, \ \ \ \Delta_2 = y^2 - z^2, \ \ \ x^2 + y^2 + z^2 = 1.
\label{Deltaidefa}
\end{equation}

Again, it can be shown that 2 of these equations, e.g.\ (\ref{3Gddelta1a}) 
and (\ref{3Gddelta2a}), are dependent on the others and can be omitted, 
leaving only 6 equations for 8 unknowns, meaning thus that the vacuum 
has a 2-dimensional degeneracy as anticipated.  Further, the equations 
(\ref{3GdDelta1})--(\ref{3Gddelta2}) imply that the quantities $\Delta_1, 
\Delta_2, \delta_1, \delta_2$ are all of order $\zeta_W^2/\zeta_S^2$, and 
that their variations to that order will give rise already to changes in 
$\theta$ and $\gamma$ of order unity.  In other words, all the properties 
surmised earlier by qualitative arguments are again confirmed.

The equations (\ref{3GdDelta1})--(\ref{3Gddelta2}) are solved
explicitly to leading order
in 
terms of $\Delta_1, \Delta_2$ and $R$ as follows:
\begin{eqnarray}
\sin \theta & = & \sqrt{\frac{2 R - 2 \Delta_1 - \Delta_2}{3 R}}\,,
\nonumber  \\
\sin \gamma & = & \sqrt{\frac{R - \Delta_1 + \Delta_2}{2 R - 2 \Delta_1
    - \Delta_2}}\,, \nonumber \\
\delta_1 & = & \sqrt{(R + 2 \Delta_1 + \Delta_2)(2 R - 2 \Delta_1 - 
\Delta_2)}\,, \nonumber \\
\delta_2 & = & \sqrt{(R - \Delta_1 - 2 \Delta_2)(R - \Delta_1 + \Delta_2)}\,,
\label{3Gsolvac}
\end{eqnarray}
or alternatively in terms of $\theta, \gamma$ and $R$ as:
\begin{eqnarray}
\Delta_1 & = & R(\cos^2 \theta - \sin^2 \theta \sin^2 \gamma), \nonumber\\
\Delta_2 & = & - R \sin^2 \theta \cos 2 \gamma,\nonumber \\
\delta_1 & = & \frac{3}{2} R \sin 2 \theta, \nonumber \\
\delta_2 & = & \frac{3}{2} R \sin^2 \theta \sin 2 \gamma.
\label{3Gsolvaca}
\end{eqnarray}
Substitution of (\ref{3Gsolvac}) into the other equations then gives $\zeta_W,
\zeta_S$, and hence $R$ also, which again all turn out to be constant,
i.e.\  
independent of $\Delta_1$ and $\Delta_2$, or of $\theta$ and $\gamma$, 
but the explicit expressions are not of immediate interest.
Similar remarks to those above for the 2-G model on the dependence of the 
vacuum on the direction of $\balpha$ apply also to the present case.  As 
$\balpha$ varies over the unit sphere, the framon triad gets distorted in
various ways from orthonormality, but the distortions are never larger than
of order $R$. 

The remarkable property of the vacuum here being degenerate and depending
on the vector $\balpha$ is of quite crucial significance to our explanation 
later for the fermion mass hierarchy and mixing pattern, and deserves a closer 
examination on its origin.  We recall that the potential $V_S$ for the strong 
framon sector by itself has a vacuum which is rather featureless where the 
strong framons just form an orthogonal triad all having the same length.  
Nevertheless, this vacuum is degenerate because of $\widetilde{su}(3)$ 
invariance, meaning that one can change at will the orientation of the 
orthogonal triad in $\widetilde{su}(3)$ space, although this degeneracy is not 
of much interest since there is no reference point to tell whether one 
has indeed made such a change.  The vector $\balpha$ coming from the weak
sector, however, gives a reference point for the orientation and distorts 
at the same time the vacua from orthonormality via the $\nu_2$ term in the 
linkage part $V_{WS}$ of the framon potential, so that the vacua now look 
different from one another depending on the value of $\balpha$.  One can
say that it is the weak sector $su(2)$ which breaks the $\widetilde{su}(3)$ 
invariance of the strong sector, in a manner similar in spirit though 
not in details to the breaking by $u(1)$ of the weak sector symmetry 
$\widetilde{su}(2)$ in the electroweak theory \cite{tHooft1,framevec}.  
Nevertheless, the theory overall has still $\widetilde{su}(3)$ invariance, 
and so must also the set of degenerate vacua labelled by $\balpha$.  And
since $\balpha$ varies under $\widetilde{su}(3)$ transformation, it has to
follow that the different vacua labelled by different values of $\balpha$ 
are equivalent under $\widetilde{su}(3)$ transformations.

Let us work this out explicitly, again for simplicity first in the 2-G model.
Let us start in a fixed $\widetilde{su}(2)$ gauge with $\balpha = \balpha_0$,
and $\balpha_0^\dagger = (1, 0)$.  This corresponds to $\alpha = 0$ in the 
notation of (\ref{2Gsolvaca}) and gives the vacuum in the triangular (local
su(2)) gauge for this value of $\balpha$ as:
\begin{equation}
\Phi_{VAC}(\balpha_0) = \left( \begin{array}{cc} 
\sqrt{\frac{1 + R}{2}} & 0 \\
0 & \sqrt{\frac{1 - R}{2}} \end{array} \right).
\label{Phivaca0}
\end{equation}
Effect now an $\widetilde{su}(2)$ transformation, say,
\begin{equation}
A = \left( \begin{array}{cc} 
   \cos \alpha e^{- i \beta} & - \sin \alpha e^{i \gamma} \\
   \sin \alpha e^{- i \gamma} & \cos \alpha e^{i \beta} \end{array} \right)
\label{Amatrix}
\end{equation}
operating on $\balpha_0$ from the left to give:
\begin{equation}
\balpha = A \balpha_0 = \left( \begin{array}{c}
   \cos \alpha e^{- i \beta} \\ \sin \alpha e^{- i \gamma} \end{array} \right).
\label{Aalpha0}
\end{equation}
The value of $\Phi$ for the vacuum corresponding to this new value of 
$\balpha$, but still in the same $\widetilde{su}(2)$ and $su(2)$ gauge, is 
obtained by operating with $A^{-1}$ from the right on $\Phi_{VAC}(\alpha_0)$ 
above, giving:
\begin{equation}
\Phi_{VAC}(\balpha) = \Phi_{VAC}(\balpha_0) A^{-1} = \left( \begin{array}{cc}
   \sqrt{\frac{1 + R}{2}} \cos \alpha e^{i \beta} &
    \sqrt{\frac{1 + R}{2}} \sin \alpha e^{i \gamma} \\
   -  \sqrt{\frac{1 - R}{2}} \sin \alpha e^{- i \gamma} &
    \sqrt{\frac{1 - R}{2}} \cos \alpha e^{- i \beta} 
   \end{array} \right).
\label{PhiVACa}
\end{equation}
This is not in the triangular form, but can be transformed into such by an
$su(2)$ transformation (i.e.\ a change of the local $su(2)$ gauge) say:
\begin{equation}
\Theta = \left( \begin{array}{cc}
   \cos \theta e^{- i \phi} & - \sin \theta e^{i \chi} \\
   \sin \theta e^{- i \chi} & \cos \theta e^{i \phi} \end{array} \right),
\label{Theta}
\end{equation}
operating from the left, thus:
\begin{equation}
\widehat{\Phi}_{VAC}(\balpha) = \Theta \Phi_{VAC}(\balpha).
\label{PhiVAChat}
\end{equation}
A straightforward calculation then shows that $\widehat{\Phi}_{VAC}(\balpha)$ 
will be triangular if we take $\phi = \beta$, $\chi = \gamma$, and
\begin{equation}
\sqrt{1 + R} \sin \theta \cos \alpha = \sqrt{1 - R} \cos \theta \sin \alpha,
\label{tantheta}
\end{equation}
or
\begin{equation}
\sin \theta = \sqrt{\frac{1 - R}{1 + R \cos 2 \alpha}} \,\sin \alpha.
\label{sintheta}
\end{equation}
Substituting this into (\ref{PhiVAChat}) gives:
\begin{equation}
\widehat{\Phi}_{VAC}(\balpha) = \left( \begin{array}{cl}
    \sqrt{\frac{1 + R \cos 2 \alpha}{2}} &
    \frac{R \sin 2 \alpha}{\sqrt{2(1 + R \cos 2 \alpha)}}
     e^{- i \beta + i \gamma} \\ 
\ \\
   0 &  \sqrt{\frac{1 - R^2}{2(1 + R \cos 2 \alpha)}}
   \end{array} \right).
\label{PhiVAChata}
\end{equation}
Comparing this with (\ref{2Gtriang}) through (\ref{2Gsolvaca}) then shows
that the two expressions are identical.  This means that the various vacua
labelled by different values of $\balpha$ in (\ref{2Gsolvaca}) are indeed
gauge equivalent as anticipated.  

A similar conclusion should hold also for the actual 3-G case.  To
check that it is so, we work again to leading order in $R$.
As in the 2-G model above, we work in the $\widetilde{su}(3)$ gauge with
$\balpha_0^\dagger = (1, 0, 0)$ and in the $su(3)$ gauge where the value 
of $\phi$ at the vacuum there is triangular, which means, according to
(\ref{3Gsolvaca}), that it is actually diagonal:
\begin{equation}
\Phi_{VAC}(\balpha_0) = \left( \begin{array}{ccc} \sqrt{\frac{1 + 2R}{3}} 
   & 0 & 0 \\  0 & \sqrt{\frac{1 - R}{3}} & 0 \\ 0 & 0 & \sqrt{\frac{1 - R}
   {3}} \end{array} \right).
\label{3GPhivaca0}
\end{equation}
A general vacuum in the same gauge is obtained by an $\widetilde{su}(3)$
transformation $A^{-1}$ operating on $\Phi_{VAC}(\balpha_0)$ from the 
right, where $A$ can as usual be parametrized as:
\begin{eqnarray}
A & = & \left( \begin{array}{ccc} e^{i \alpha_1} & 0 & 0 \\ 0 & e^{i \alpha_2} 
   & 0 \\ 0 & 0 & e^{- i \alpha_1 - i \alpha_2} \end{array} \right)
    \left( \begin{array}{ccc} c_3 & - s_3 e^{-i \sigma_3} & 0 \\
      s_3 e^{i \sigma_3} & c_3 & 0 \\ 0 & 0 & 1 \end{array} \right)
     \nonumber \\  &   & \times
    \left( \begin{array}{ccc} c_2 & 0 & - s_2 e^{-i \sigma_2} \\
      0 & 1 & 0 \\ s_2 e^{i \sigma_2} & 0 & c_2 \end{array} \right) 
    \left( \begin{array}{ccc} 1 & 0 & 0 \\ 0 & c_1 & - s_1 e^{-i \sigma_1} \\
      0 & s_1 e^{i \sigma_1} & c_1 \end{array} \right),
\label{Ageneral}
\end{eqnarray}
where we have abbreviated sines and cosines of various angles to $s_i, c_i$.
Although the 3-G vacuum is labelled by these $A$ and not just by the
vector $\balpha$ as in the 2-G case,
for the present calculation we restrict ourselves only to those
$A$'s with $s_1 = 0$, i.e.\ no rotation in the 23-plane.  This is sufficient
to take $\balpha_0$ to any $\balpha$, say (\ref{3Galpha}).  Following then
the notation there for the phases, we have for $A$ the expression:
\begin{equation}
A = \left( \begin{array}{ccc} c_2 c_3 e^{-i \beta_1} & - s_3 e^{-i(\beta_1
     + \sigma_3)} & - c_3 s_2 e^{i(\beta_3 + \beta_2 + \sigma_3)} \\
     c_2 s_3 e^{-i \beta_2} & c_3 e^{-i(\beta_2 + \sigma_3)} & - s_2 s_3 
     e^{i(\beta_1 + \beta_3 + \sigma_3)} \\ s_2 e^{-i \beta_3} & 0 &
     c_2 e^{i(\beta_1 + \beta_2 + \sigma_3)} \end{array} \right).
\label{Aspecial}
\end{equation}
This gives for the vacuum at $\balpha$:
\begin{eqnarray}
& & \Phi_{VAC}(\balpha) = \Phi_{VAC}(\balpha_0) A^{-1} =  \nonumber \\
\ \nonumber \\
& & \left( \begin{array}{lll}
   \sqrt{\frac{1 + 2 R}{3}} c_2 c_3 e^{i \beta_1} &
   \sqrt{\frac{1 + 2 R}{3}} c_2 s_3 e^{i \beta_2} &
   \sqrt{\frac{1 + 2 R}{3}} s_2 e^{i \beta_3} \\
\ \\
   - \sqrt{\frac{1 - R}{3}} s_3 e^{i(\beta_1 + \sigma_3)} &
   \sqrt{\frac{1 - R}{3}} c_3 e^{i(\beta_2 + \sigma_3)} & 0 \\
\ \\
   - \sqrt{\frac{1 - R}{3}} s_2 c_3 e^{-i(\beta_2 + \beta_3 + \sigma_3)} &
   - \sqrt{\frac{1 - R}{3}} s_2 s_3 e^{-i(\beta_1 + \beta_3 + \sigma_3)} &
   \sqrt{\frac{1 - R}{3}} c_2 e^{-i(\beta_1 + \beta_2 + \sigma_3)} 
   \end{array} \right). \nonumber\\
\label{3GPhiVACa}
\end{eqnarray}
$\Phi_{VAC}(\balpha)$ can be returned to the triangular form by applying
from the left an $su(3)$ transformation $\Theta$ of the same form as $A$ 
in (\ref{Aspecial}) only with $s_i, c_i$ replaced by $s_i', c_i'$, where
\begin{equation}
\sqrt{\frac{1 + 2R}{3}} \, c_2 s'_2 = \sqrt{\frac{1 - R}{3}}\,s_2 c'_2,
\label{3Gtan1}
\end{equation}
and
\begin{equation}
s_2 c_3 s'_3 = s_2' c'_3 s_3.
\label{3Gtan2}
\end{equation}
A straightforward, though somewhat lengthy, calculation via (\ref{3Gtan1}),
(\ref{3Gtan2}) and (\ref{3Gsolvaca}) then shows that the resulting triangular
form is indeed identical to (\ref{3Gtriang}).  This means that one has shown 
that also for the 3-G case the vacua for different values of $\balpha$ are
gauge equivalent to one another, although here so far only to leading order 
in $R$

We recall that all these intriguing properties of the vacuum arise from the 
special form of the framon potential which is itself a consequence of the 
double invariance under both the local $u(1) \times su(2) \times su(3)$ and 
the global $\tilde{u}(1) \times \widetilde{su}(2) \times \widetilde{su}(3)$ 
symmetries basic to the whole framon idea.  We note in particular the 
delicate interplay between the electroweak and strong sectors via the
linkage terms implied by the invariance, which distort the
strong vacuum from its symmetry position, and couple at the same time the
distorted vacua to the vector $\balpha$ coming from the weak sector, so
that if the vacuum moves, it will bring the vector $\balpha$ with it.  We
see, therefore, that the chain of logic in arriving at the above results 
is quite intricate and tightly knit.

\setcounter{equation}{0}

\section{Confinement and the Boson Spectrum}

The framon potential as given, for example, in (\ref{VPhivec}), with 
negative $|\bphi|^2$ terms is of a form usually interpreted as typical 
of spontaneously broken gauge symmetries.  But, as pointed out as long 
ago as 1978 by 't~Hooft \cite{tHooft} and by Banks and Rabinovici
\cite{Banovici}, and again as repeatedly emphasized more recently by 
't~Hooft \cite{tHooft1}, the electroweak theory which has such a potential 
can equally be interpreted as a confining theory where the gauge symmetry
$su(2)$ remains exact, what is broken being just (in our language) the 
global $\widetilde{su}(2)$ symmetry ``dual'' to it.  And the 2 pictures, 
i.e.\ whether confinement or spontaneously broken symmetry, are mathematically 
equivalent in present applications of the electroweak theory, although 
in the long run at ultra high energies, they may diverge in physical 
content.  For a discussion of the confinement picture in the context of 
the framon idea we are here exploring, see for example \cite{framevec}.

That being the case, we shall adopt in the present treatment, as our 
assumption (B), exclusively the confinement picture which we find both 
practically more convenient and conceptually more appealing, although much 
of what we do here, though perhaps not all, could equally be performed 
in the symmetry breaking picture.  This means that we shall consider both 
the $su(2)$ and the $su(3)$ local gauge symmetries as confining, with 
only the difference that the confinement in $su(2)$ is much deeper than 
in $su(3)$.  This last is a necessary assumption for the confinement 
picture to hold as a realistic description of nature since the confinement 
by colour $su(3)$ is already revealed to us by present experiment in the 
sense that we have already for decades been probing into hadrons as 
compound states of colour confinement revealing their structure in terms 
of their constituents.  On the other hand, the equivalent compound states 
in $su(2)$ confinement, such as leptons and quarks, still appear to us as 
point-like in all experiments so far performed.  It was thus assumed by 
't~Hooft and implicitly also by Banks and Rabinivici that confinement in 
$su(2)$ is considerably deeper than with colour, so that only in future 
deep inelastic experiments with ultra high energies not yet available to 
us will the structure of quarks and leptons be revealed.  We accept this 
assumption, for the moment without further question, although we hope to 
return to it with respect to a nonabelian version of electromagnetic 
duality at a later stage for reasons to be explained in the concluding
section.

In the confinement picture then, let us ask what is the boson spectrum for 
the theory that has been constructed.  To answer this question, we have 
first to make clear what exactly this question means.  The $su(2)$ and 
$su(3)$ gauge symmetries being both confined, it follows that only states 
which belong to the singlet representation in both will be observable as 
propagating particles in free space.  However, while performing a deep 
inelastic scattering experiment of an electron on a proton, for example, 
we are already probing inside the proton where coloured objects such as 
quarks and gluons can propagate freely, and only $su(2)$ remains confined.  
Let us call this the standard model scenario, for this is indeed what the 
Standard Model, as it now stands, represents in terms of the scheme under 
consideration.  In the standard model scenario then, the boson spectrum
would mean a list of all $su(2)$ singlet boson states, whether colour
neutral or otherwise, which can be formed as bound or compound states
from the fundamental fields by $su(2)$ confinement.   Given the assumption
that $su(2)$ confinement is much deeper than colour $su(3)$ confinement,
these states will appear as quasi-elementary under all present experimental 
conditions. On the other hand, one can consider also what one can call 
the soft hadron scenario where experimental conditions are such that one 
never probes inside hadrons and sees all hadrons as quasi-elementary 
objects as one did some decades ago.  Then the boson spectrum would mean 
a list of the bound or compound states which can be formed by confining 
together, via either $su(2)$ or $su(3)$ confinement or both, the fundamental 
gauge boson and framon fields so far considered and eventually also the 
fundamental fermion fields yet to be introduced. 

Start then with the standard model scenario where only the electroweak
$su(2)$ but not colour $su(3)$ confines.  Now of the fundamental boson 
fields introduced in section 2, only the weak gauge field  $B_\mu$ and 
framon field $\bphi$ carry (local) $su(2)$ indices and have to be confined; 
the others, i.e.\ the colour gauge fields $C_\mu$ and strong framon field 
$\bphi_a$, being already (local) $su(2)$ singlets, can exist as freely 
propagating particles inside hadrons.  What is then the boson spectrum in 
such a situation?

We notice that, apart from the $\nu_1, \nu_2$ terms in the framon potential 
linking the weak and strong framon sectors, the part of the action involving 
the weak gauge and framon fields $B_\mu$ and $\bphi$ in the present framework 
is identical to that of the standard electroweak theory.  Hence, an analysis 
similar to that in \cite{tHooft,Banovici,framevec} for the pure electroweak 
theory is expected to give for the boson spectrum in $su(2)$ confinement 
essentially the same result.  Explicitly, let us choose to work in the gauge
when the framon field $\bphi$ points in the up direction and is real.  In 
this gauge, the part of the action involving only the weak fields will 
reduce to the standard electroweak action in the symmetry breaking picture 
evaluated in the unitary gauge when the vacuum expectation value for $\bphi$ 
points in the up direction and is real, with only the difference that the 
$su(2)$ gauge field $B_\mu$ is everywhere replaced by its gauge transform:
\begin{equation}
\tilde{B}_\mu = \frac{i}{g_2} 
\,\Omega^\dagger (\partial_\mu - i g_2 B_\mu)\, \Omega\,,
\label{Btilde}
\end{equation}
where $\Omega$ is the transformation we have used above to fix the gauge.
Notice that $\Omega^\dagger$ being the transformation which rotates the
local $su(2)$ frame to align it with the global $\widetilde{su}(2)$
frame, $\Omega$ transforms from the left under $su(2)$ but from the right
under $\widetilde{su}(2)$, so that $\tilde{B}_\mu$ in (\ref{Btilde}), with 
all $su(2)$ indices saturated, is an $su(2)$ singlet.  Indeed, to leading 
order in the fluctuations $H$ of $|\bphi| = F + H$ about its vacuum value 
$F = \sqrt{\mu_W/ 2 \lambda_W}$ at the minimum of the potential $V_W$, we 
can write:
\begin{equation}
   \Tr\,(\Phi^\dagger (\partial_\mu - i g_2 B_\mu) 
\Phi) \approx ig_2 \tilde{B}_\mu + \cdots \,,
\label{Btildea}
\end{equation}
where $\Phi$ is the $2 \times 6$ matrix representing the ``weak'' framon 
field  $\phi_r^{\tilde{r} \tilde{a}}$, and the trace Tr is taken over 
only the $\widetilde{su}(3)$ indices.  We see thus that the massive vector
boson $\tilde{B}_\mu$, which in the conventional treatment is 
considered to be 
the gauge boson having acquired a mass via symmetry breaking, is now to be 
regarded as a $p$-wave bound state of the ``weak'' framon with its own 
conjugate.  Similarly, it can be seen from:
  \begin{equation}
\tr \left( \Tr (\Phi^\dagger \Phi) \right) \approx F^2 + 2 FH + \cdots \,,
\label{HTrtr}
\end{equation}
where the 2 traces are taken over $\tilde{r}$ and $\tilde{a}$ indices,
that the standard Higgs boson $H$ is now to be regarded as the $s$-wave 
bound state of the framon-anti-framon pair.  But since the action is the 
same as in the pure electroweak theory in the symmetry breaking picture 
except for these reinterpretations, we recover the same mass spectrum with 
the familiar mixing between $\gamma$ and $Z$.  Indeed, apart from the 
trace over $\tilde{a}$ indices, these arguments are very similar to those
in the pure electroweak theory, some more details of which can 
be found e.g.\ in \cite{framevec}.

There is one difference, however, with the conventional electroweak theory 
in that because the ``weak'' framons here carry a $\tilde{u}(1)$ charge, 
so also will their bound states.  This can easily be worked out for 
the various states just by summing the charges of the constituents and 
are listed below together with the $u(1)$ charge as $[q, \tilde{q}]$:
\begin{equation}
\begin{array}{rclcl}
H & \sim & \Tr\,(\Phi^{\dagger} \Phi)\,, & \quad& [0, 0]\,;\\
W_\mu^+ & \sim & \bphi^{(-) \dagger} (A_\mu^i \tau_i)\,\bphi^{(+)}, &
   \quad & [1, -1]\,;  \\
W_\mu^- & \sim & \bphi^{(+) \dagger} (A_\mu^i \tau_i)\,\bphi^{(-)}, &
   \quad & [- 1, 1]\,;  \\
\gamma{\rm -}Z & \sim & \bphi^{(\pm) \dagger} 
(A_\mu^i \tau_i)\,\bphi^{(\pm)}, &
   \quad & [0, 0]\,. 
\label{bosonqqt}
\end{array}
\end{equation}
The question of what this $\tilde{u}(1)$ charge $\tilde{q}$ signifies
will be postponed until fermions are introduced in the next section

Apart from this small but, as we shall see, important, difference of a new
conserved quantum number $\tilde{q}$, the result obtained above in the 
standard model scenario for the purely weak sector is otherwise seen to 
be identical to the conventional electroweak theory.  Together with the 
colour gauge boson $C_\mu$, i.e.\ the gluon, which in the standard model 
scenario remains massless, these make up exactly the boson content of the 
Standard Model.  Hence, we conclude that in the standard model scenario, 
the present scheme differs in the boson content from the Standard Model 
only in the strong framons $\bphi_a$.  Thus, so long as these last do not 
spoil the usual comparison of QCD with experiment as, for example, in the 
running of the coupling $\alpha_s$, which we have hope to show later, for
the reason mentioned in the concluding section, to be indeed the case, we 
can claim consistency.  The $\nu_1, \nu_2$ terms in the framon potential 
so far neglected will mix the Higgs boson $H$ with hadron states in a 
manner to be made explicit later, which will change somewhat the properties 
of $H$ and its mass but not the boson content in the standard model scenario.

The above conclusion on the boson content in the standard model scenario 
is obtained by essentially just a paraphrase in our more complex framework 
of 't~Hooft's and Banks and Rabinovici's original arguments which took 
account only of the lowest bound states formed from a framon-antiframon 
pair via $su(2)$ confinement.  However, if we were to take seriously 
the confinement picture not just as a convenience, or alternative to the 
conventional symmetry-breaking picture for the electroweak theory, but as 
actually physical, then it would be difficult to exclude the possibility
of there being other boson states formed by $su(2)$ confinement, such as
radial and orbital excitations of the ground states, familiar to hadron 
spectroscopists in the parallel scenario of colour $su(3)$ confinement.  
We have then to envisage a situation where the actual boson spectrum is 
considerably richer than that seen at present, where the additional states, 
being presumably much heavier than the ground states listed, are not 
noticeable at present, but will be revealed later when higher experimental
energies become available.  Such a situation will obtain not just in our
special framework, but in any model where the confinement picture for the
electroweak theory is taken as physical \cite{Abfarhi1,Abfarhi2,CFJ,CFrit}, 
and has been analysed in this context extensively and in detail already in 
the early paper by Claudson, Farhi and Jaffe \cite{CFJ}.  
They not only listed 
the many additional states possible, some with quite exotic features, 
but also examined the effects these can have on the physics accessible 
to present energies, for example those on the electromagnetic form factors 
of quarks and leptons.  Apart from other results, their analysis on the 
latter effects confirms the intuition one had that, provided the additional 
states are sufficiently heavy, they would not disturb the correspondence 
between the confinement and symmetry-breaking pictures proposed.  However, 
this still leaves the question why the ground vector boson states $W$ and 
$Z-\gamma$ are so much lighter than their radial and orbital excitations, 
in seeming contrast to what is seen in the hadron spectrum obtained in the 
parallel case of colour $su(3)$ confinement.  At one's present stage of 
understanding of $su(2)$ confinement, unfortunately, one is still in no 
better position to answer this question than in the days of \cite{CFJ}.  
We note only at the speculative level an interesting possibility, to be 
discussed in the concluding section of this paper, which is connected to 
a nonabelian generalization of electromagnetic duality and to the question 
mentioned above of why $su(2)$ confinement is supposedly so much deeper than 
colour confinement.  Such questions, however, of whether the confinement 
picture can indeed be taken physical, and if so, what new effects or 
phenomena it will imply at higher energies than that presently available, 
can be postponed as they are not of immediate concern
for our present limited purpose here of just deducing from the 
constructed framework some otherwise mystifying properties of the Standard 
Model. For this purpose, the ground state boson spectrum listed above for 
the standard model scenario is already sufficient.  

Next, what about the soft hadron scenario when colour is also confined? 
In that case, neither the gluon $C_\mu$ nor the framon $\bphi_a$ will 
remain free but have to be confined with one another or with coloured 
fermions into colour singlets to give a host of new states to the hadron 
spectrum.  The details of this is not of immediate interest in the present
context of attempting to construct a protogenic standard model. The hadron 
spectrum is already so complicated in the Standard Model and so little 
understood that some more states will not be easily detected or missed, 
especially if they are heavy and unstable as these new modes promise to be. 
In any case, we are certainly not yet at the stage to worry about such 
details.  However, for reasons to be made clear later related to the 
linkage between the weak and strong sectors via the $\nu_1, \nu_2$ terms, 
we have particular interest in the $s$-wave hadron bound states via colour
confinement of ${\bphi}_a^{\dagger} \bphi_a$, namely the equivalents in 
the strong sector of the Higgs field $H$ in the electroweak sector, for 
which we wish now to work out the spectrum.

For the strong framon potential $V_S$ by itself, this was done already in
\cite{framevec}.  We wish now, however, to take account also of the linkage
$\nu_1, \nu_2$ terms which distort the strong vacuum from orthonormality
and make the calculation considerably more complicated.  Nevertheless, the
tactics developed there can still be applied with modifications.  To do so, 
we choose again to work in the triangular gauge, where we recall that the 
strong framon field $\bphi_a$ or $\Phi$ at vacuum is parametrized as in 
(\ref{3Gtriang}).  We are now to consider small fluctuations about the 
vacuum, which we represent as:
\begin{equation}
\Phi = \left( \begin{array}{ccc}
  X \cos \delta_1 + h_1 & (X \sin \delta_1 \sin \gamma + \eta_3)\, e^{i \chi_3}
      & (X \sin \delta_1 \cos \gamma + \eta_2)\, e^{i \chi_2} \\
    0 & Y \cos \delta_2 + h_2 & (Y \sin \delta_2 + \eta_1)\, e^{i \chi_1} \\
    0 & 0 & Z + h_3 \end{array} \right),
\label{3Gtriangfl}
\end{equation} 
where $h_1, h_2, h_3$ are real and $\eta_1, \eta_2, \eta_3$ complex.  These
fluctuations give altogether 9 degrees of freedom, representing the 9 real
(strong) Higgs fields formed as bound states of $\Phi^{\dagger} \Phi$.  To 
readers more familiar with the symmetry breaking picture, it may help to 
add that the 8 zero modes appearing in that picture which are to be eaten 
up by the colour gauge bosons are here already taken out of consideration 
by fixing to the triangular gauge.

The fluctuations $h_i$ and $\eta_i$ span the Higgs modes but are as yet
neither mass eigenstates nor do they form
an orthonormal basis.  Again to avoid
complicated details clouding the logic, we shall work these out first
in the 2-G model, for which the parallel for (\ref{3Gtriangfl}) is:
\begin{equation} 
\Phi = \left( \begin{array}{cc}
    X \cos \delta + h_1 & (X \sin \delta + \eta)\, e^{i \phi} \\
    0 & Y + h_2 \end{array} \right).
\label{2Gtriangfl}
\end{equation}
First, we need to be explicit in the unitary transformation taking the 
original frame in which the vacuum value of $\Phi$ is triangular and the 
fluctuations appear as:
\begin{equation}
\Phi = \left( \begin{array}{cc} 
   X \cos \delta + \delta \phi_1^{\tilde{1}} & (X \sin \delta 
   + \delta \phi_1^{\tilde{2}}) e^{i \phi} \\  
   \delta \phi_2^{\tilde{1}} & Y + \delta \phi_2^{\tilde{2}} 
   \end{array} \right),
\label{2Gorigfl}
\end{equation}
to the frame where $\Phi$ itself including fluctuations is of the triangular
form (\ref{2Gtriangfl}).  This is readily seen to first order in the 
fluctuations to be:
\begin{equation}
\Omega_{DF} = \left( \begin{array}{cc}
    1 - \frac {i}{X \cos \delta}\, \delta \phi^{\tilde{1}}_{1 I} &
    \frac{1}{X \cos \delta} (\delta \phi^{\tilde{1}}_2)^* e^{i \phi} \\
\ \\
    - \frac{1}{X \cos \delta}\, \delta \phi^{\tilde{1}}_2\, e^{- i \phi} &
    1 + \frac{i}{X \cos \delta}\, \delta \phi^{\tilde{1}}_{1 I},
    \end{array} \right)\,,
\label{OmegaDF}
\end{equation}
where the subscript $I$ denotes the imaginary part.  Multiplying 
$\Omega_{DF}$ 
to (\ref{2Gorigfl}) and comparing with (\ref{2Gtriangfl}) then gives:
\begin{eqnarray}
h_1 & = & \delta \phi^{\tilde{1}}_{1 R} \nonumber \\
h_2 & = & \delta \phi^{\tilde{2}}_{2 R}
    - (\tan \delta)\, \delta \phi^{\tilde{1}}_{2 R} \nonumber \\
\eta_R & = & \delta \phi^{\tilde{2}}_{1 R}
    + \frac{y}{x \cos \delta}\, \delta \phi^{\tilde{1}}_{2 R} \nonumber \\
\eta_I & = & \delta \phi^{\tilde{2}}_{1 I}
    - \frac{y}{x \cos \delta}\, \delta \phi^{\tilde{1}}_{2 I}
    -  (\tan \delta)\, \delta \phi^{\tilde{1}}_{1 I},
\label{hindeltaphi}
\end{eqnarray}
where $R$ denotes the real part.  We can consider then the coefficients 
of $\delta \phi^{\tilde{a}}_{aR}$ and $\delta \phi^{\tilde{a}}_{aI}$ as 
an 8-vector.  The vectors representing $h_1, h_2, \eta_R, \eta_I$ above
are not all normalized nor all mutually orthogonal as 8-vectors.  But we 
can construct easily from these via the standard Gram-Schmidt procedure 
an orthonormal basis for the 4-dimensional subspace they span, say $H_1, 
H_2, H_R, H_I$, which are conveniently represented as $2 \times 2$ complex
matrices, giving respectively:
\begin{eqnarray}
\widehat{V}_1 & = & 
\left( \begin{array}{cc} 1 & 0 \\ 0 & 0 \end{array} \right)
    \nonumber \\
\widehat{V}_2 & = & \sqrt{x^2 \cos^2 \delta + y^2}
\left( \begin{array}{cc} 0 & \frac{xy \sin \delta}{x^2
       \cos^2 \delta +y^2}\, e^{i \phi} \nonumber\\
\ \nonumber \\
    - \frac{x^2 \sin \delta \cos \delta}{x^2 \cos^2 \delta + y^2}
\, e^{- i \phi} & 1 \end{array} \right) \nonumber \\
\widehat{V}_R & = & \frac{1}{\sqrt{x^2 \cos^2 \delta + y^2}}
\left( \begin{array}{cc} 0 & x \cos \delta\, e^{i \phi} \\
    y\, e^{- i \phi} & 0 \end{array} \right) \nonumber \\
\ \nonumber\\
\widehat{V}_I & = & i \left( \begin{array}{cc} -x \sin \delta  & x \cos 
\delta\, e^{i \phi} \\
    - y\, e^{- i \phi} & 0 \end{array} \right).
\label{2GV}
\end{eqnarray}

The mass matrix of these Higgs states $H_K$ can be worked out by substituting
(\ref{2Gtriangfl}) into the expression for the framon potential and expanding
to second order in the fluctuations $h_1, h_2, \eta_R, \eta_I$ and $H$ (the 
fluctuation of the weak framon $\phi$ about its vacuum value), obtaining 
a quadratic form in these.  One then re-expresses this form in terms of the
orthonormal basis $H_1, H_2, H_R, H_I$ listed above.  The coefficients of
the various terms then give the Higgs mass matrix, which is a little involved. 
Since we shall not need this in detail, we shall give the answer only to 
first order in our supposedly small parameter $\zeta_W/\zeta_S$:
\begin{equation}
M_H = \left( \begin{array}{ccccc}
    0 & \zeta_W \zeta_S (2 \nu_1 - \nu_2) & \zeta_W \zeta_S \nu_2 \cos 2 \alpha
    & - \zeta_W \zeta_S \nu_2 \sin 2 \alpha & 0 \\
\ \\
    {*} & 2(2 \lambda_S + \kappa_S) \zeta_S^2 & - 2 (\lambda_S + \kappa_S)
    \zeta_S^2 \Delta & 2 (\lambda_S + \kappa_S) \zeta_S^2 \delta & 0 \\
\ \\
    {*} & * & 2 \kappa_S \zeta_S^2 & 0 & 0 \\
\ \\
    {*} & * & * & 2 \kappa_S \zeta_S^2 & 0 \\
\ \\
    {*} & * & * & * & 2 \kappa_S \zeta_S^2 \end{array} \right),
\label{MHiggs}
\end{equation}
where the rows and columns are labelled, in order, by $H, H_+, H_-, H_R, H_I$
with $H_{\pm} = (1/\sqrt{2}) (\pm H_1 + H_2)$.  We note only that the mass
eigenstates will mix the ``real'' states $H, H_+, H_-, H_R$ but not the
``imaginary'' state $H_I$, the mixing being of order $\zeta_W/\zeta_S$, or
lower, so that it can in principle be worked out by perturbation theory to 
that order. For what follows, however, we shall not need the explicit form
of the mass eigenstates.

The calculation for the actual 3-G case starting from (\ref{3Gtriangfl})
follows along the same lines but the algebra is more complicated.  
We give here the transformation 
$\Omega_{DF}$ to first order in the fluctuations $\delta 
\phi^{\tilde{a}}_a$:
\begin{equation} \Omega_{DF} = \left(
\begin{array}{ccc}
\Omega_{11} & \Omega_{12} & \Omega_{13} \\
\Omega_{21} & \Omega_{22} & \Omega_{23} \\
  \Omega_{31} & \Omega_{32} & \Omega_{33} \end{array} \right),
\label{OmegaDF3G}
\end{equation}
where
\begin{eqnarray}
\Omega_{11} &=& 1 - \frac{i}{X \cos \delta_1} (\delta
\phi_1^{\tilde{1}})_I \nonumber\\
\Omega_{12} &=& \frac{(\delta \phi_2^{\tilde{1}})^* e^{i \chi_3}}{X \cos
     \delta_1} \nonumber \\
\Omega_{13} &=& \frac{(\delta \phi_3^{\tilde{1}})^* e^{i \chi_2}}{X \cos
     \delta_1} \nonumber \\
\Omega_{21} &=& - \frac{\delta \phi_2^{\tilde{1}} \, e^{-i \chi_3}}{X \cos
     \delta_1} \nonumber \\
\Omega_{22} &=& 1 - \frac{i (\delta \phi_2^{\tilde{2}})_I}{Y \cos
   \delta_2} + \frac{i \tan \delta_1 \sin \gamma (\delta
   \phi_2^{\tilde{1}})_I}{Y \cos \delta_2} \nonumber \\
\Omega_{23} &=& \frac{(\delta \phi_3^{\tilde{2}})^* e^{i \chi_1}}{Y
   \cos \delta_2} - \frac{\tan \delta_1 \sin \gamma (\delta
   \phi_3^{\tilde{1}})^* e^{i \chi_1}}{Y
   \cos \delta_2} \nonumber \\
\Omega_{31} &=& - \frac{\delta
   \phi_3^{\tilde{1}} \, e^{-i \chi_2}}{X \cos \delta_1} \nonumber \\
\Omega_{32} &=& - \frac{\delta \phi_3^{\tilde{2}} e^{-i \chi_1}}{Y \cos
   \delta_2} + \frac{\tan \delta_1 \sin \gamma \delta
   \phi_3^{\tilde{1}} e^{-i \chi_1}}{ Y \cos \delta_2} \nonumber \\
\Omega_{33} &=& 1 - \frac{i (\delta \phi_3^{\tilde{3}})_I}{Z} +
\frac{i \tan \delta_1 \cos \gamma (\delta \phi_3^{\tilde{1}})_I}{Z}
+ \frac{i \tan \delta_2 (\delta \phi_3^{\tilde{2}})_I}{Z} \nonumber \\
&& - \frac{i
   \tan \delta_1 \tan \delta_2 \sin \gamma (\delta
   \phi_3^{\tilde{1}})_I}{Z} 
\end{eqnarray}
and to first order in $R$
the 9 orthonormal Higgs states in the analogous matrix notation of 
(\ref{2GV}) which will be of use later:
\begin{eqnarray}
\widehat{V}_1 
& = & \left( \begin{array}{ccc} 1 & 0 & 0 \\ 0 & 0 & 0 \\ 0 & 0 & 0
    \end{array} \right) \nonumber \\
\ \nonumber\\
\widehat{V}_2 & = & \left( \begin{array}{ccc}
    0 & \frac{\delta_1 x y \sin \gamma}{x^2 + y^2}\, e^{i \chi_3} & 0
    \\
\ \\
    -\frac{\delta_1 x^2 \sin \gamma}{x^2 + y^2}\, e^{- i \chi_3} & 1 &
    0 \\
\  \\
    0 & 0 & 0 \end{array} \right) \nonumber \\
\ \nonumber\\
\widehat{V}_3 & = & \left( \begin{array}{ccc}
    0 & 0 & \frac{\delta_1 x z \cos \gamma}{x^2 + z^2}\, e^{i \chi_2}
    \\ 
\ \\
    0 & 0 & \frac{\delta_2 y z}{y^2 + z^2}\, e^{i \chi_1} \\
\ \\
    -\frac{\delta_1 x^2 \cos \gamma}{x^2 + z^2}\, e^{- i \chi_2} &
      -\frac{\delta_2 y^2} {y^2 + z^2}\, e^{- i \chi_1} & 1
    \end{array} \right) \nonumber \\
\ \nonumber\\
\widehat{W}_1^R & = & \frac{1}{\sqrt{y^2 + z^2}} \left( \begin{array}{ccc}
    0 & \frac{\delta_1 x y^2 \cos \gamma}{x^2 + y^2}\, e^{i \chi_3} &
      \frac{\delta_1 x z^2 \sin \gamma}{x^2 + z^2}\, e^{i \chi_2} \\
\ \\
    -\frac{\delta_1 x^2 y \cos \gamma}{x^2 + y^2}\, e^{- i \chi_3} &
      0 & y e^{i \chi_1} \\
\ \\
    -\frac{\delta_1 x^2 z \sin \gamma}{x^2 + z^2}\, e^{- i \chi_2} &
      z e^{- i \chi_1} & 0 \end{array} \right) \nonumber \\
\ \nonumber\\
\widehat{W}_2^R & = & \frac{1}{\sqrt{x^2 + z^2}} \left( \begin{array}{ccc}
    0 & -\frac{\delta_2 x y^2}{x^2 + y^2}\, e^{i \chi_3} & x e^{i
      \chi_2} \\
\ \\
    \frac{\delta_2 x^2 y}{x^2 + y^2} e^{- i \chi_3} & 0 & 0 \\
\ \\
    z e^{- i \chi_2} & 0 & 0 \end{array} \right) \nonumber \\
\ \nonumber\\
\widehat{W}_3^R & = & \frac{1}{\sqrt{x^2 + y^2}} \left( \begin{array}{ccc}
    0 & x e^{i \chi_3} & 0 \\
    y e^{- i \chi_3} & 0 & 0 \\
    0 & 0 & 0 \end{array} \right) \nonumber \\
\ \nonumber\\
\widehat{W}_1^I & = & \frac{i}{\sqrt{y^2 + z^2}} \left( \begin{array}{ccc}
    0 & -\frac{\delta_1 x y^2 \cos \gamma}{x^2 + y^2}\, e^{i \chi_3} &
      \frac{\delta_1 x z^2 \sin \gamma}{x^2 + z^2}\, e^{i \chi_2} \\
\ \\
   - \frac{\delta_1 x^2 y \cos \gamma}{x^2 + y^2}\, e^{- i \chi_3} &
      - \delta_2 y & y e^{i \chi_1} \\
\ \\
    \frac{\delta_1 z x^2 \sin \gamma}{x^2 + z^2}\, e^{- i \chi_2} &
      - z e^{- i \chi_1} & 0 \end{array} \right) \nonumber \\
\ \nonumber\\
\widehat{W}_2^I & = & \frac{i}{\sqrt{x^2 + z^2}} \left( \begin{array}{ccc}
    - x \delta_1 \cos \gamma & -\frac{\delta_2 x y^2}{x^2 + y^2}\, e^{i \chi_3}
      & x e^{i \chi_2} \\
\ \\
    -\frac{\delta_2 x^2 y}{x^2 + y^2}\, e^{-i \chi_3} & 0 & 0 \\
\ \\
    - z e^{- i \chi_2} & 0 & 0 \end{array} \right) \nonumber \\
\ \nonumber\\
\widehat{W}_3^I & = & \frac{i}{\sqrt{x^2 + y^2}} \left( \begin{array}{ccc}
    - x \delta_1 \sin \gamma & x e^{i \chi_3} & 0 \\
    - y e^{- i \chi_3} & 0 & 0 \\
    0 & 0 & 0 \end{array} \right).
\label{3GV}
\end{eqnarray}

\setcounter{equation}{0}

\section{The Fermion Sector}

Extending next our consideration to the fermion sector, we have first to
specify what we are to take as our fundamental fermion fields, which are,
of course, to be given as representations of the gauge symmetries of the 
theory, namely $u(1) \times su(2) \times su(3)$.  Not having ascribed in
our framework any geometrical significance to fermion fields as we did to 
the vector and scalar fields in the boson sector, we are thrown back as
usual to arguments of simplicity.  On this guideline then, let us take as
fundamental fermion fields the following:
\begin{equation}
\psi = \psi(1, 1); \ \ \ \psi_r = \psi(2, 1); \ \ \ 
\psi_a = \psi(1, 3); \ \ \ \psi_{ra} = \psi(2, 3),
\label{psis}
\end{equation}
where inside the brackets, the first argument denotes the dimension
of the representation of the $su(2)$ symmetry, the second argument that of 
the $su(3)$ symmetry.  These are the simplest possibilities involving as they 
do only the singlet and fundamental representations of either $su(2)$ or 
$su(3)$.

We have yet to specify the $u(1)$ charges $q$ of the $\psi$'s in 
(\ref{psis}),
if any.  Recall now the discussion in section 2 on the choice of the gauge 
group, which concluded with the choice of the group $U(1,2,3)$ defined there. 
The admissible $u(1)$ charges for the fermions (\ref{psis}) which are of 
necessity representations of the group $U(1,2,3)$ can then be read off from
(\ref{qadmit}).  Let us concentrate first on the $su(2)$ doublets $\psi(2, 1)$ 
and $\psi(2, 3)$ and restrict ourselves as usual to those with the smallest 
values of $|q|$, obtaining:
\begin{eqnarray}
& \psi^{({\pm 1/2})}_r = \psi^{({\pm 1/2})}(2, 1); \ \ \ & q = \pm \half, \\
& \psi^{({1/6})}_{ra} = \psi^{({1/6})}(2, 3); \ \ \ & q = \tfrac{1}{6},
\label{fermiondq}
\end{eqnarray}
where when necessary, the $u(1)$ charge $q$ will be carried as a bracketed 
superscript as shown.

Notice that the fundamental fermion fields, like the gauge boson fields, 
do not, and have no reason to, carry any indices referring to the global 
symmetries $\tilde{u}(1) \times \widetilde{su}(2) \times \widetilde{su}(3)$. 
These have to do only with the global reference frames and should thus 
affect only the framons.

In the confinement picture in which we work, only singlets of the nonabelian 
local symmetries can exist as freely propagating particles.  Let us ask then 
what fermion spectrum we are to expect.  Again, the answer depends on whether 
we are in the standard model or soft hadron scenario.  Our primary concern 
is the standard model scenario where experimental conditions are at present, 
namely where colour $su(3)$ confinement has already been revealed, but 
$su(2)$ confinement is not, so that only $su(2)$ singlets are observed as 
freely propagating and appear to us as elementary objects.  In that case, 
only $\psi(1, 1)$ and $\psi(1, 3)$ in (\ref{psis}) can appear free but 
$\psi(2, 1)$ and $\psi(2, 3)$ have to be confined.

As in the pure electroweak theory treated by 't~Hooft \cite{tHooft} and 
Banks and Rabinovici \cite{Banovici} (see also \cite{framevec}), the lowest 
fermion bound states obtained from $\psi(2, 1)$ and $\psi(2, 3)$ by $su(2)$
confinement are expected from their binding with those scalar framon fields
carrying $su(2)$ indices, namely the ``weak'' framons $\phi_r^{\tilde{r} 
\tilde{a}}$, thus: 
\begin{eqnarray}
\chi^{\tilde{r} \tilde{a}} & = & \sum_r (\phi_r^{\tilde{r} \tilde{a}})^*
    \psi_r; \label{chiss} \\
\chi_a^{\tilde{r} \tilde{a}} & = & \sum_r (\phi_r^{\tilde{r} \tilde{a}})^*
    \psi_{ra}.
\label{chis}
\end{eqnarray}
These are identified in the electroweak theory as respectively left-handed
leptons and quarks, which assignment we here also adopt.

If we take the confinement picture as actually physical, and not merely as 
an alternative presentation of the conventional symmetry-breaking picture
in the electroweak theory, then, just as in the case of boson bound states 
formed from framons considered before, there can in principle be many 
other fermion states obtainable as excitations of leptons and quarks from 
binding the fundamental fermions $\psi(2, 1), \psi(2, 3)$ with framons via 
$su(2)$ confinement.  At the moment, however, for our immediate purpose 
of deducing Standard Model properties from the constructed framework, we 
shall be concerned only with the leptons and quarks given explicitly above 
as the lowest fermion bound states.

The first thing of interest to note in the above expressions is that the
bound states, in contrast to the fundamental fermions $\psi$'s, now carry
global indices, $\tilde{r}$, $\tilde{a}$ and indeed also $\tilde{q}$ not
yet exhibited, which they acquire through their framon constituents.  In
particular, the $\tilde{r}$ indices for the global $\widetilde{su}(2)$
symmetry, which occurred already in the pure electroweak theory treated by,
for example 't~Hooft \cite{tHooft1}, was taken there as representing up-down
flavour.  Explicitly, for the present scheme, if we take $\tilde{r} = \mp$, 
we would have, with $\psi_r = \psi_r^{(-1/2)}$ in (\ref{chiss}),
just by adding up the charges from their constituents, the 
following $u(1)$ and $\tilde{u}(1)$ charges $[q, \tilde{q}]$ for the bound 
states $\chi$,
\begin{equation}
\left( \begin{array}{c} \chi^{(-) \tilde{a}} \\  \chi^{(+) \tilde{a}}
    \end{array} \right) \sim \left(\!\!\!\! \begin{array}{ccc} 
{}& [0, -\half]&{} \\ {}& [-1, \half]&{}
    \end{array} \!\!\!\!\right); \ \ 
\left( \begin{array}{c} \chi_a^{(-) \tilde{a}} \\  \chi_a^{(+) \tilde{a}}
    \end{array} \right) \sim \left(\!\!\!\! \begin{array}{ccc} 
{} &[\frac{2}{3}, -\half] &{} \\{}& [-\third, \half] &{}
    \end{array} \!\!\!\!\right).
\label{chiqqt}
\end{equation}
We see that we obtain in the first doublet (of $\widetilde{su}(2)$) exactly 
the right $u(1)$ charges for the leptons $(\nu, e^-)$, and for the second 
doublet those for the quarks $(u, d)$.

What is new, however, is that the leptons and quarks have now acquired, in 
addition to flavour, a charge $\tilde{q}$ for the global $\tilde{u}(1)$ as 
well as an index $\tilde{a}$ referring to the global $\widetilde{su}(3)$ 
symmetry.  As already indicated in the introduction, assumption (C), we 
would associate the $\widetilde{su}(3)$ symmetry with fermion generations, 
a theme which we shall develop further in the rest of the present paper. 
For the moment, let us first ask what is the physical meaning, if any, of 
the $\tilde{u}(1)$ symmetry and charge.

The theory having been constructed to be invariant under both $u(1)$ and 
$\tilde{u}(1)$ by virtue of the framon hypothesis, it follows that both
the $u(1)$ charge $q$ and the $\tilde{u}(1)$ charge $\tilde{q}$ are conserved
quantities.  This ought then to imply another conserved quantity in addition 
to the electric charge $q$.  However, when applied to framons where the 
$\tilde{u}(1)$ charge originates, it gives nothing new, since $\tilde{q}$ 
there is always equal in magnitude and opposite in sign to the electric charge 
$q$ so that the conservation of one necessarily implies the conservation of 
the other.  But such a situation will not in general be maintained when 
applied to compound particles formed from framons with other fields, which 
carry only the charge $q$, not the charge $\tilde{q}$.  An examination of 
(\ref{bosonqqt}), 
however, shows that for the boson states $H, W^\pm$, and 
$\gamma-Z$, the condition $\tilde{q} = - q$ still obtains so that again no 
new conserved quantity appears.  This is because these boson states are 
formed from a framon-antiframon pair, some with a gauge boson $B_\mu$ which 
carries neither charge.  But with leptons and quarks formed from a fundamental 
charged fermion with a single framon field, the condition $\tilde{q} = - q$ 
no longer applies.  Instead, as seen in (6.6), the value of $\tilde{q}$ is 
shifted from $-q$ as follows:
\begin{equation}
\tilde{q} = - q + \frac{1}{2} (B - L),
\label{BminusL}
\end{equation}
where the special amount of the shift can be traced eventually to the
charge assignments to the fundamental fermion fields, which in turn arise 
from the choice of $U(1,2,3)$ as gauge group.  That being the case, the 
conservation of both $q$ and $\tilde{q}$ as required by the theory now 
implies not just the conservation of the electric charge $q$ but also the 
conservation of the baryon-lepton number $B - L$.

It thus seems that the new global $\tilde{u}(1)$ invariance obtained through 
implementing the framon idea has indeed a physical meaning, and a rather
important one, namely baryon-lepton number conservation. Now baryon number 
conservation, or in its modern form as $B - L$ conservation, is a mystery 
which has puzzled us particle physicists for some time.  As long ago as the 
1950's, Lee and Yang \cite{LeeYang} already asked the searching question 
why we knew of no gauge invariance principle which implies baryon number 
conservation, although it is one of the best kept conservation law known 
in nature.  It is thus very interesting that the framon idea under 
consideration provides now an answer for it.  In the framon scheme, baryon 
number (or $B - L$ in modern terms) is ``dual'' to the electric charge in 
a similar sense as flavour $\widetilde{su}(2)$ is ``dual'' to the confining 
local gauge symmetry $su(2)$, or, as generations is ``dual'' to $su(3)$ 
colour, as we hope next to show.

As already noted, the compound fields $\chi$ in (\ref{chis}) carry also an 
$\tilde{a}$ index acquired from their framon constituents which means in 
physical terms that our left-handed leptons and quarks should each exist 
in $\widetilde{su}(3)$ (anti-)triplets.  Since the theory is invariant under 
$\widetilde{su}(3)$ one would need in any case to account for the existence 
of this multiplet structure as one did for the $\tilde{u}(1)$ charge.  It 
seems thus natural to try associating it with fermion generations, for which 
there has been wanting a plausible explanation already for decades.

To see whether such a supposition might work, let us try first to construct
Yukawa couplings for the fermions in (\ref{psis}) with the ``weak'' framon 
fields.  In analogy to the standard electroweak theory, we suggest the 
following:
\begin{eqnarray}
{\cal A}_{\rm YK} &=& \sum_{[\tilde{a}] [b]} Y^{\rm lepton}_{[b]} 
\bar{\psi}^r_{[\tilde{a}]}
    \phi_{r}^{(-) \tilde{a}} \half (1 + \gamma_5) \psi^{[b]}
    + \sum_{[\tilde{a}] [b]} Y'^{\rm lepton}_{[b]} \bar{\psi}^r_{[\tilde{a}]}
    \phi_{r}^{(+) \tilde{a}} \half (1 + \gamma_5) \psi'^{[b]}
    \nonumber \\
&& {} + {\rm h.c.}
\label{Yukawal}
\end{eqnarray}
for leptons, and
\begin{eqnarray}
{\cal A}_{\rm YK} &=& \sum_{[\tilde{a}] [b]} Y^{\rm quark}_{[b]} 
\bar{\psi}^{ra}_{[\tilde{a}]}
    \phi_{r}^{(-) \tilde{a}} \half (1 + \gamma_5) \psi_a^{[b]}
    +  \sum_{[\tilde{a}] [b]} Y'^{\rm quark}_{[b]} 
\bar{\psi}^{ra}_{[\tilde{a}]}
    \phi_{r}^{(+) \tilde{a}} \half (1 + \gamma_5) \psi_a^{'[b]}
    \nonumber \\
&& {} + {\rm h.c.}
\label{Yukawaq}
\end{eqnarray}
for quarks, where the indices in brackets $[\tilde a]$ and $[b]$ label just 
3 identical copies of the same fields.  These expressions are by construction
invariant under Lorentz transformations and the internal symmetries $su(2)$
and $su(3)$.  Invariance under $\widetilde{su}(2)$ can be seen by recalling
from (\ref{su2tphi}) the expressions of $\phi_r^{(\pm)}$ as the inner products 
$\balpha^{(\pm)}\cdot\bphi_r$ 
between 2 $\widetilde{su}(2)$ vectors.  What remains to
be checked is then just the invariance with respect to $u(1), \tilde{u}(1)$ and
$\widetilde{su}(3)$.  Invariance under $u(1)$ can be guaranteed by assigning
appropriate $u(1)$ charges to the right-handed fields, thus $q = 0$ for 
$\psi^{[b]}$, $q = -1$ for $\psi'^{[b]}$, $q = 2/3$ for $\psi_a^{[b]}$, and 
$q = - 1/3$ for $\psi_a^{'[b]}$, namely the same charges as the 
corresponding compound state $\chi$ in each case.  This procedure is the 
same as in the conventional formulation of the Standard Model.  We only 
need to check here that these charge assignments to the right-handed 
fields are consistent with those allowed by the gauge group $U(1,2,3)$ 
as listed in (\ref{qadmit}), and one sees that they are.  Similarly, 
$\tilde{u}(1)$ invariance can be ensured also by assigning to the 
right-handed fields the same $\tilde{u}(1)$ charges $\tilde{q}$ as their 
corresponding compound states.  This contravenes in principle our rule 
that fundamental fermions should not carry global charges or indices.  
This is however just a matter of convenience, for these charges being 
global can equally be absorbed into the Yukawa couplings $Y_{[b]}$ etc.  
We follow the standard convention and assign the $\tilde{u}(1)$ charge to 
the right-handed field so that they carry the same baryon and lepton number 
as their left-handed partners and so guarantee $\tilde{u}(1)$ invariance.  
There remains then only $\widetilde{su}(3)$ invariance, of which, strictly 
speaking, the expressions (\ref{Yukawal}) and (\ref{Yukawaq}) do not have.  
Under an $\widetilde{su}(3)$ transformation, the framons $\bphi$ transform
but the left-handed fermions $\psi^{[\tilde{a}]}$ with which the $\tilde{a}$
index of the framons are contracted, do not transform.  However, if we force
a transformation on these latter fields, it would mean only a relabelling of
these identical fields, and physics should not be affected.  Thus
$\widetilde{su}(3)$ invariance is maintained in this more specialized sense.  
This is similar to the argument used below, by relabelling the right-handed 
fields, to cast the fermion mass matrix into a hermitian form  often found 
useful in the usual Standard Model.  We thus propose to accept the above 
Yukawa couplings as valid on this basis.

Suppose we do, then on substituting the vacuum expectation value $\zeta_W$ 
of the weak framon field into the Yukawa couplings above, one obtains mass 
matrices of the form:
\begin{equation}
m \sim \left( \begin{array}{c} \alpha^{\tilde{1}} \\ \alpha^{\tilde{2}} \\
    \alpha^{\tilde{3}} \end{array} \right) (Y_{[1]}, Y_{[2]}, Y_{[3]})
    \half (1 + \gamma_5)
    + \left( \begin{array}{c} Y^*_{[1]} \\ Y^*_{[2]} \\ Y^*_{[3]} \end{array}
    \right) (\alpha^{\tilde{1}*}, \alpha^{\tilde{2}*}, \alpha^{\tilde{3}*})
    \half (1 - \gamma_5).
\label{massmat}
\end{equation}
Equivalently, by relabelling the right-handed fields appropriately, we can
recast the mass matrix in a hermitian form with no $\gamma_5$ \cite{Weinberg}
as:
\begin{equation}
m = m_T \left( \begin{array}{c} \alpha^{\tilde{1}} \\ \alpha^{\tilde{2}} \\
    \alpha^{\tilde{3}} \end{array} \right)
    (\alpha^{\tilde{1}*}, \alpha^{\tilde{2}*}, \alpha^{\tilde{3}*}).
\label{massmatW}
\end{equation}
In other words, we obtain that the mass matrix of all quarks and leptons
are similar, being just a product of the same vector $\balpha$ in generation 
space by its hermitian conjugate, and differ for different fermion species, 
i.e.\ whether lepton or quark or whether flavour up or down, only by the 
normalization:
\begin{equation}
m_T = \zeta_W \rho_T^2,
\label{mT}
\end{equation}
with
\begin{equation}
\rho_T^2 = |Y_{[1]}|^2 + |Y_{[2]}|^2 + |Y_{[3]}|^2.
\label{rhoT}
\end{equation}
This form for the fermion mass matrix plays a very significant role in our 
understanding of the fermion mass hierarchy and mixing pattern observed 
in experiment, a theme that will be taken up again in the next section.

Before we do so, let us first make a necessary detour to consider the soft 
hadron scenario where colour $su(3)$ as well confines.  As remarked before, we 
are not at present interested in details of the hadronic spectrum and will not 
therefore examine in general terms the many fermion states which are obtained 
by confining via colour the fundamental fermion fields introduced above with 
the scalar framon fields.  We shall only, for a reason which will be made
clear later, work out a special case as example.  For our purpose, it will 
be sufficient to consider the (hadronic) fermion states obtained by confining 
$\psi_a$ via colour confinement with the ``strong'' framon from which we 
shall even omit the $\tilde{r}$ index as being here inessential and write 
it as $\phi_a^{\tilde{a}}$.

Our first task is to construct a Yukawa coupling term for $\psi_a$ with
$\phi_a^{\tilde{a}}$.  We notice that whereas the framon field carries an
$\tilde{a}$ index for $\widetilde{su}(3)$, the fermion fields do not.  To
maintain $\widetilde{su}(3)$ invariance, therefore, we need here a vector 
in $\widetilde{su}(3)$ space to saturate this $\tilde{a}$ index.  As noted
already in \cite{framevec}, there is no such vector available to play this
role within the purely strong sector.  But in the present set-up, there
is the vector $\balpha$ coming from the weak sector which can possibly be 
so employed.  In thus introducing a vector originating from the weak sector 
so as to construct a Yukawa term in the strong sector, one is imitating, 
in spirit though not in detail, the previous construction of the Yukawa 
terms in the weak sector (\ref{Yukawal}) and (\ref{Yukawaq}) by means of 
the $\widetilde{su}(2)$ vectors $\balpha^{(\pm)}$ originating from the 
electromagnetic $u(1)$ sector so as to guarantee $\widetilde{su}(2)$
invariance, and in that case, one obtained the standard electroweak theory
result.  The vector $\balpha$, however, does not have a definite value,
but, the vacuum being degenerate as asserted previously, can point in any 
direction in $\widetilde{su}(3)$ space.  Nevertheless, these directions 
being all gauge equivalent, it should not matter which particular value we 
choose.  Let us then just select an arbitrary direction, say, $\balpha_0$ 
and construct our Yukawa term as follows:
\begin{equation}
{\cal A}_{\rm YK} = \sum_{[b]} Z_{[b]} \bar{\psi}^a \bphi_a
\cdot \balpha_0
    \frac{1}{2}(1 + \gamma_5) \psi^{[b]} + {\rm h.c.},
\label{Yukawas}
\end{equation}
where $\psi^{[b]},\ b = 1, 2, 3$ are again just 3 copies of right-handed 
fields.  This has then all the invariance properties that are required 
and seems to be the only Yukawa term with these properties that one can 
construct in the strong sector from the quantities so far introduced.  
What the chosen value for $\balpha_0$ represents will become clear later 
as the formalism develops.  

Again, for consequences of such a Yukawa term, because of the distortion 
of the strong framon vacuum from orthonormality, the working out is a little 
complicated, and will be given here in detail only for the 2-G model.  In
that case, for any $\balpha_0$ that we may have chosen, the theory being 
invariant under $\widetilde{su}(2)$, we can always choose to work in the
gauge where $\balpha_0$ points in the up direction and is real, thus:
\begin{equation}
\balpha_0 = \left( \begin{array}{c} 1 \\ 0 \end{array} \right).
\label{alpha0fixed}
\end{equation}
Further, the theory being also invariant locally under $su(2)$, we can also
choose to work in the local $su(2)$ gauge where the vacuum value for the
framon field $\Phi$ corresponding to this $\balpha_0$ is triangular, which 
in this case actually means diagonal, as given in (\ref{Phivaca0}) above. 
We shall refer henceforth to this chosen gauge as the $T_0$-gauge.

Our next task is to evaluate the mass matrix for the fermion states formed
from $\psi_a$ and $\phi_a^{\tilde{a}}$ by colour confinement.  As for the
leptons and quarks in the weak sector, the mass matrix for these is obtained
by substituting for $\Phi$ in the Yukawa term its vacuum value.  For the
general vacuum labelled by $\balpha$, the vacuum value for $\Phi$ is given
already in (\ref{2Gtriang}), with $\delta$ and $\Delta$ given in terms of 
$\alpha$ in (\ref{2Gsolvaca}).  This is, however, in the $su(2)$ gauge, say 
the $T$-gauge, where the vacuum value of $\Phi$ is triangular, and this gauge
changes from vacuum to vacuum labelled by different values of $\balpha$.
In what follows we shall be interested in comparing the mass matrix, and 
hence the vacuum values of $\Phi$ at different $\balpha$'s and this will 
make sense only if we keep the same gauge, let us say the $T_0$-gauge
defined above.  The value of $\Phi$ at the vacuum labelled by
$\balpha$
in the $T_0$ gauge
can be obtained either by operating from the left on (\ref{2Gtriang})
with
$\Theta^{-1}$ for 
$\Theta$ in (\ref{Theta}), changing thus from the original $T$-gauge to the
$T_0$-gauge we prefer, or else by operating from the right on (\ref{Phivaca0})
with $A^{-1}$ for 
$A$ in (\ref{Amatrix}), i.e.\ translating the vacuum value at the 
reference point $\balpha_0$ to the general point $\balpha$.  In either case,
one obtains for the vacuum value of $\Phi$ at $\balpha$ as (\ref{PhiVACa}),
which we rewrite here again for future convenience:  
\begin{equation}
V_0(\balpha) = \Phi_{VAC}(\balpha_0) A^{-1} = \left( \begin{array}{cc}
    \sqrt{\frac{1 + R}{2}} \cos \alpha e^{i \beta} &
    \sqrt{\frac{1 + R}{2}} \sin \alpha e^{i \gamma} \\
   -  \sqrt{\frac{1 - R}{2}} \sin \alpha e^{- i \gamma} &
    \sqrt{\frac{1 - R}{2}} \cos \alpha e^{- i \beta} 
   \end{array} \right).
\label{V0}
\end{equation}

Substituting the vacuum value (\ref{V0}) of $\Phi$ into (\ref{Yukawas}),
one obtains then for the fermion bound states via $su(2)$ 
(2-G colour) confinement
formed from $\psi_a$ with $\phi_a^{\tilde{a}}$ the mass matrix:
\begin{equation}
{\bf m} = \zeta_s \,(V_0 \balpha_0)\, (Z_{[1]}, Z_{[2]}) \frac{1}{2}(1
+ \gamma_5)
     + \left( \begin{array}{c} Z_{[1]}^* \\ Z_{[2]}^* \end{array} \right)
      (V_0 \balpha_0)^\dagger \frac{1}{2}(1 - \gamma_5),  
\label{massmatH}
\end{equation}
where we have used a bold-faced letter ${\bf m}$ to remind ourselves that 
we are dealing here with hadron states and to distinguish it from the mass 
matrix $m$ introduced above for leptons and quarks.  Also, to save on the
notation, we have omitted from $V_0$ its argument, leaving its dependence
on $\balpha$ understood, as we shall do as well for related quantities in 
future.  Again, as was done before for (\ref{massmat}), by appropriately 
relabelling the right-handed singlet fermions $\psi^{[b]}$, we can rewrite 
${\bf m}$ in the hermitian form independent of $\gamma_5$ as:
\begin{equation}
{\bf m} = {\bf m}_T |v_0 \rangle \langle v_0|,
\label{massmatHW}
\end{equation}
with 
\begin{equation}
|v_0 \rangle = V_0 \balpha_0.
\label{v0ket}
\end{equation}
Explicitly,
\begin{equation}
{\bf m}_T = \zeta_s \rho_S; \ \ \ \rho_S = \sqrt{|Z_{[1]}|^2 + |Z_{[2]}|^2},
\label{rhoS}
\end{equation}
and:
\begin{equation}
|v_0 \rangle 
   = \left( \begin{array}{c}  \sqrt{\frac{1 + R}{2}} \cos \alpha
     e^{i \beta} \\ - \sqrt{\frac{1 - R}{2}} \sin \alpha e^{- i \gamma}
     \end{array} \right).
\label{v0keta}
\end{equation}

For our future use, we need also to derive the couplings of these bounds
state fermions to the Higgs bosons listed in the preceding section.  These
couplings are obtained by expanding the Yukawa coupling term (\ref{Yukawas})
to first order in fluctuations of the framon fields about their vacuum 
values.  With right-handed fermions labelled as in (\ref{massmatHW}), these 
couplings for the fluctuations $\delta \phi_a^{\tilde{a}}$ are easily seen
to be given just as:
\begin{eqnarray}
\Gamma_{aR}^{\tilde{a}} & = & \rho_S |v \rangle \langle v_0| 
   \frac{1}{2}(1 + \gamma_5) + \rho_S |v_0 \rangle \langle v| 
   \frac{1}{2}(1 - \gamma_5) \nonumber \\
\Gamma_{aI}^{\tilde{a}} & = & i \rho_S |v \rangle \langle v_0| 
   \frac{1}{2}(1 + \gamma_5) - i \rho_S |v_0 \rangle \langle v|
   \frac{1}{2}(1 - \gamma_5),
\label{Gammaaat}
\end{eqnarray}
with:
\begin{equation}
|v \rangle = V \balpha_0; \ \ \ 
   (V)_b^{\tilde{b}} = \delta_{ab} \delta^{\tilde{a} \tilde{b}}.
\label{vket}
\end{equation}

What we need, however, are the couplings for the Higgs states $H_1$, 
$H_2$, 
$H_R$, $H_I$ identified in the preceding section.  These were given as linear 
combinations of $\delta \phi_a^{\tilde{a}}$ in (\ref{hindeltaphi}) in terms of 
the matrices $\widehat{V}_K$. But these were in the $T$-gauge which we have 
first to transform back into the selected $T_0$-gauge, thus:
\begin{equation}
V_K = \Theta^{-1} \widehat{V}_K.
\label{2GVK}
\end{equation}
The corresponding coupling of the $K$th Higgs state to the fermion states
is then given as::
\begin{equation}
\Gamma_K = \rho_S |v_K \rangle \langle v_0| \frac{1}{2}(1 + \gamma_5) 
   + \rho_S |v_0 \rangle \langle v_K| \frac{1}{2}(1 - \gamma_5),
\label{GammaK}
\end{equation}
for $K = 1, 2, R, I$, where for $V_K$ listed as in (\ref{2GVK}):
\begin{eqnarray}
|v_1 \rangle & = & V_1 \balpha_0 = \left( \begin{array}{c}
   \cos \theta e^{i \beta} \\ - \sin \theta e^{-i \gamma}
   \end{array} \right), \nonumber \\
|v_2 \rangle & = & V_2 \balpha_0 = \frac{1}{\sqrt{x^2 \cos^2 \delta + y^2}}
   \left( \begin{array}{c} - x^2 \sin \theta \sin \delta \cos \delta 
   e^{i \beta}  \\  - x^2 \cos \theta \sin \delta \cos \delta e^{-i \gamma} 
   \end{array} \right), \nonumber \\
|v_R \rangle & = & V_R \balpha_0 = \frac{1}{\sqrt{x^2 \cos^2 \delta + y^2}}
   \left( \begin{array}{c} y \sin \theta e^{i \beta} \\
   y \cos \theta e^{-i \gamma} \end{array} \right), \nonumber \\
|v_I \rangle & = & V_I \balpha_0 = i \left( \begin{array}{c} 
   (- x \cos \theta \sin \delta - y \sin \theta) e^{i \beta}  \\
   ( x \sin \theta \sin \delta - y \cos \theta) e^{-i \gamma}
   \end{array} \right).
\label{vKket}
\end{eqnarray}

There is no difficulty of principle extending the above considerations to
the actual 3-G situation.  For the present, however, since the solution
for the vacuum has only been found to leading order in the supposedly
small parameter $R$, and the gauge relationship between them only worked
out for a special class of $\widetilde{su}(3)$ transformations $A$, the 
same restrictions remain in what follows.  Apart from these restrictions,
the formulae (\ref{massmatHW})---(\ref{v0ket}) above for the mass matrix 
of the analogous bound state remain essentially the same, only with the 
matrix $V_0$ being now the 3-G version given in (\ref{3GPhiVACa}).  The 
couplings of these fermion bound states to the Higgs fields also remain 
basically the same as in (\ref{GammaK}) although notation-wise we find it 
convenient to distinguish the 3 types of Higgs $K = k, kR, kI, k = 1, 2, 3$ 
thus:
\begin{eqnarray}
\Gamma_k & = & \rho_S |v_k \rangle \langle v_0| \frac{1}{2}(1 + \gamma_5) 
   + \rho_S |v_0 \rangle \langle v_k| \frac{1}{2}(1 - \gamma_5), \\
\Gamma_k^R & = & \rho_S |w_k^R \rangle \langle v_0| \frac{1}{2}(1 + \gamma_5) 
   + \rho_S |v_0 \rangle \langle w_k^R| \frac{1}{2}(1 - \gamma_5), \\
\Gamma_k^I & = & \rho_S |w_k^I \rangle \langle v_0| \frac{1}{2}(1 + \gamma_5) 
   + \rho_S |v_0 \rangle \langle w_k^I| \frac{1}{2}(1 - \gamma_5),
\label{GammakrI}
\end{eqnarray}
where:
\begin{equation}
|v_k \rangle = \Theta^{-1} \widehat{V}_k \balpha_0; \ \ \ 
|w_k^R \rangle = \Theta^{-1} \widehat{W}_k^R \balpha_0; \ \ \ 
|w_k^I \rangle = \Theta^{-1} \widehat{W}_k^I \balpha_0,
\label{vkRIket}
\end{equation}
for $\widehat{V}_k, \widehat{W}_k^R, \widehat{W}_k^I$ listed in 
(\ref{3GV}).  The 
resulting formulae for these vectors are a little complicated 
even to first order in $R$.  Since the algebra is straightforward, we shall
give the answer here only for $|v_0 \rangle$:
\begin{equation}
|v_o \rangle = \left( \begin{array}{l}
  \sqrt{\frac{1 + 2R}{3}}\, c_2 c_3\, e^{i \beta_1} \\
  - \sqrt{\frac{1 - R}{3}} \,s_3\, e^{i(\beta_1 + \sigma_3)} \\
  - \sqrt{\frac{1 - R}{3}} \,s_2 c_3\, e^{-i(\beta_2 + \beta_3 + \sigma_3)}
  \end{array} \right)\;,
\label{v0ket3}
\end{equation}
which is just the first column of the matrix $V_0$ in (\ref{3GPhiVACa})
and will be of use later.

\setcounter{equation}{0}

\section{The Rotating Fermion Mass Matrix}

By a rotating mass matrix here we mean one which changes its orientation 
in generation space with changing scale, in much the same way as a running
coupling changes with changing scale its value.  A coupling changes with
scale because of renormalization effects; so in the same way a mass matrix
may change its orientation in generation space.  It is in fact quite easy 
to imagine or construct situations in which it does do so.

The reason we are interested in a rotating fermion mass matrix is that it
has been suggested that this may lead to mass hierarchy and mixing patterns
for fermions similar to what have been observed in experiment.  Indeed, one 
can even claim there is circumstantial evidence that the observed mass and 
mixing patterns for leptons and quarks are due to a rotating fermion mass 
matrix \cite{cevidsm}.  Furthermore, a phenomenological model (DSM) has been
constructed which gives a good description of these phenomena, including
in particular neutrino oscillations, depending on only a small number of
parameters \cite{phenodsm,genmixdsm}.

Starting anew from more considered premises, we have now constructed, to
us, a much more theoretically attractive and internally consistent model.
We naturally wish to ask whether in this new model the fermion mass matrix
will still rotate and give rise to a similar explanation for the mass and
mixing patterns.  At first sight, this might seem difficult, since the
fermions we are interested in are the leptons and quarks which have rather
little structure in our new model, being bound states via the deeply 
confining $su(2)$ symmetry, and should appear at present experimental
conditions as approximate point particles.  Although they are made to 
carry a generation index in view of the weak framon field which is one
of their constituents and which carries this index, this index is only global
and brings with it no new interactions.  Indeed, this is in a sense
fortunate for one would not want, and probably cannot admit, any blatant 
new interactions for leptons and quarks for these may spoil the already 
good description of these particles by the Standard Model.  For this reason, 
in the lepton and quark mass matrix given above in (\ref{massmatW}), the 
generation index appeared only in the global factor $\balpha$ which has 
no explicit coupling to any gauge field.

However, the interesting thing is that this vector $\balpha$ is coupled
nevertheless to the strong sector via the framon potential constructed to 
satisfy the symmetries intrinsic to the system.  Indeed, as shown above
in section 4, the vacuum of the framon potential is degenerate and depends
on the direction of $\balpha$ so that if the vacuum moves from one value 
to another within the degenerate set, which it can do under renormalization
effects in the strong sector, then $\balpha$ can change in direction,
i.e.\ 
in other words, rotate.  The attention is then shifted to the behaviour of 
the vacuum under renormalization, which we shall now study.

The vacuum value of the framon field $\Phi$, we noticed above, occurs in 
the mass matrix ${\bf m}$ for the hadron bound state in (\ref{massmatHW}) 
via (\ref{v0ket}).  We can thus obtain information on the change in value
of the vacuum value of $\Phi$, and hence on the rotation of $\balpha$, by 
studying the behaviour of ${\bf m}$ in (\ref{massmatHW}) under renormalization.
Now a similar problem has been studied before in the phenomenological model 
DSM \cite{ckm}, where it was found that the rotation, if any, would come 
mostly from insertions of Higgs loops in the fermion propagator (i.e.\ for 
the hadronic bound state fermions, in the present language), the effect of 
other loops being suppressed.  It is for this reason that we have worked 
out above in some detail the spectra of these fermion and Higgs bound states 
as well as their couplings.  With these results, we can now calculate the 
effect of Higgs loop insertions, which we shall do below to 1-loop order. 

The insertion of a Higgs loop to the fermion self-energy is of the form:
\begin{equation}
\Sigma(p) = \frac{i}{(4 \pi)^4} \sum_K \int d^4 k \frac{1}{k^2 - M_K^2}
    \Gamma_K \frac{(p\llap/ - k\llap/) + \m}{(p - k)^2 - {\m}^2} \Gamma_K,
\label{Sigma}
\end{equation}
where we may for the moment take $K$ to label the Higgs mass eigenstates.
After standard manipulations, regularizing the divergence by dimensional
regularization, one obtains: 
\begin{equation}
\Sigma(p) = - \frac{1}{16 \pi^2} \sum_K \int_0^1 dx \Gamma_K
    \{\bar{C} - \ln (Q^2/\mu^2) \} [p\llap/ (1 - x) + \m]\, \Gamma_K,
\label{Sigma2}
\end{equation}
where
\begin{equation}
Q^2 = \m^2 x + M_K^2 (1 - x) - p^2 x(1 -x),
\label{Qsquare}
\end{equation}
with $\bar{C}$ being the divergent constant to be subtracted in the 
standard $\overline{\rm MS}$ scheme.  The renormalization to the mass 
matrix $\delta {\bf m}$ is obtained by first commuting the $p\llap/$ in the 
numerator half to the extreme left and half to the extreme right, then putting 
$p\llap/ = {\bf m}$ and $p^2 = {\bf m}^2$.  The full explicit expression 
for $\delta \m$ so obtained together with more details of the calculation 
can be found in \cite{transmudsm}.  Here, we are interested only in the 
terms dependent on the scale $\mu$.  These are of two types.  First, 
there are terms of the form:
\begin{equation}
\Gamma_K \m \Gamma_K = \rho_S^2 \langle v_0|v_0 \rangle
    \langle v_0|v_K \rangle |v_K \rangle \langle v_0|
    \half(1 + \gamma_5) + {\rm c.c.}
\label{GammamGamma}
\end{equation}
Then there are terms of the form:
\begin{eqnarray}
\lefteqn{ \Gamma_K p\llap/ \Gamma_K  \rightarrow} \nonumber \\ 
&&\half \rho_S^2
    \{\langle v_0|v_0 \rangle \langle v_K|v_0 \rangle |v_K \rangle
    \langle v_0| + \langle v_0| v_0 \rangle \langle v_K|v_K \rangle
    |v_0 \rangle \langle v_0| \} \half (1 + \gamma_5) \nonumber \\
  && + \ {\rm c.c.}
\label{GammapGamma}
\end{eqnarray}
In (\ref{GammamGamma}) and (\ref{GammapGamma}), we have already commuted 
$p\llap/$ to the left and right as stipulated and used the known forms
for the tree-level mass matrix (\ref{massmatHW}) and Higgs couplings
(\ref{GammaK}).

We notice that the second term in (\ref{GammapGamma}) is proportional to 
the tree level mass matrix (\ref{massmatHW}), and will thus only change 
its normalization, not its orientation.  Any rotation will thus have to 
come from the other terms, either of which will add a correction of the 
form:
\begin{equation}
\sum_K \langle v_0|v_K \rangle |v_K \rangle,
\label{gov1}
\end{equation}
or
\begin{equation}
\sum_K \langle v_K|v_0 \rangle |v_K \rangle,
\label{gov2}
\end{equation}
to the vector $|v_0 \rangle$ on the left, and will change the orientation 
of the mass matrix, provided of course that these corrections are neither
vanishing nor parallel to $|v_0 \rangle$ itself.  For this reason, we call
(\ref{gov1}) and (\ref{gov2}) the governing vectors for rotation, and it 
is on these now that our attention is turned.

Again, to avoid getting bogged down by algebraic complications too soon,
we shall work out first the governing vectors in the 2-G model.  We recall
that the index $K$ is meant here to label the Higgs mass eigenstates, not
the same as in (\ref{vKket}).  This presents no difficulty in principle, 
for we had the Higgs mass matrix in (\ref{MHiggs}), which if necessary 
can be diagonalized to give the mass eigenstates as orthogonal transforms
of the vectors in (\ref{2GV}), from which eigenstates, the corresponding 
$|v_K \rangle$ can be recalculated.  However, this will not be necessary 
since an orthogonal transformation among the vectors $V_K$ will leave 
the governing vectors invariant.  We can thus evaluate the governing vectors 
with just the old set of $|v_K \rangle$ in (\ref{vKket}).  The answer is as 
follows.  First, we notice by (\ref{tantheta}) or (\ref{sintheta}) that
both $|v_2 \rangle$ and $|v_R \rangle$ are orthogonal to $|v_0 \rangle$ giving
thus no contributions, while $|v_1 \rangle$ is parallel to $|v_0 \rangle$.
Hence, summing over the ``real'' $K$'s, we have:
\begin{equation}
\sum_{K = 1,2, R} \langle v_0|v_K \rangle |v_K \rangle
   = \sum_{K = 1,2, R} \langle v_K|v_0 \rangle |v_K \rangle
   = |v_0 \rangle.
\label{govR2G}
\end{equation} 
This means that the ``real'' states will give no rotation to the mass
matrix.  The ``imaginary'' state $K = I$, on the other hand, gives the 
following contribution:
\begin{equation}
|{\rm gvI} \rangle = \langle v_0|v_I \rangle |v_I \rangle
   = - \langle v_I|v_0 \rangle |v_I \rangle
   = - \frac{1}{2} R \sin 2 \alpha \left( \begin{array}{c}
     \sqrt{\frac{1 + R}{2}} \sin \alpha e^{i \beta} \\
     \sqrt{\frac{1 - R}{2}} \cos \alpha e^{-i \gamma} \end{array} \right),
\label{govI2G}
\end{equation}
where we have used the relations (\ref{tantheta}) and (\ref{2Gsolvaca})
to simplify the expression to the given form.  This being in general 
neither vanishing nor parallel to $|v_0 \rangle$, we conclude that in the 
2-G case, the vector $|v_0 \rangle$ will indeed in general rotate with 
changing scale as we hope, and that by section 4 above it will drag the 
vector $\balpha$ along with it.   

Indeed, substituting the result (\ref{govI2G}) into (\ref{Sigma2}), one 
obtains the following renormalization group equation for the rotating 
vector $|v_0 \rangle$:
\begin{equation}
\frac{d}{d(\ln \mu^2)} |v_0 \rangle = - \frac{3}{64 \pi^2} \rho_S^2 
   |{\rm gvI} \rangle,
\label{rge}
\end{equation}
where one has neglected terms which changes only the normalization of
the vector which is not of immediate interest here.  The equation makes
the rotation effect with respect to scale change explicit.  In particular, 
one notes that at $\alpha = n \pi/2$, the governing vector vanishes.  This 
means that these are fixed points of the rotation which will be of special
significance later when practical applications to fermion mixing are 
considered. 

To visualize more easily the behaviour of the rotating vectors between 
fixed points, let us simplify the equation by taking the leading order
in an expansion in the supposedly small parameter $R$, in which case we
have:
\begin{equation}    
|{\rm gvI} \rangle \sim -\frac{1}{2} R \sin 2 \alpha \left( \begin{array}{c}
   \sin \alpha e^{i \beta} \\ \cos \alpha e^{-i \gamma} \end{array} \right),
\label{gvIapprox}
\end{equation}
of first order in smallness and orthogonal to $|v_0 \rangle$.  Suppose we
start with $\balpha$ in the first quadrant, i.e.\  $0 < \alpha < \pi/2$,
which gives then, according to (\ref{v0ket}) a $|v_0 \rangle$ in the 4th 
quadrant.  Then, according to the rotation equation (\ref{rge}) and the
expression (\ref{gvIapprox}) for $|{\rm gvI} \rangle$, a change in the scale 
$\mu$ by a positive amount will give a vector increment to $|v_0 \rangle$ 
pointing back in the first quadrant.  This means that for increasing $\mu$,
$|v_0 \rangle$ will rotate in the counterclockwise direction, or that
$\balpha$ will rotate in the clockwise direction.  Hence, on increasing 
$\mu$ further, $\alpha$ will eventually reach the limiting value $0$, or
$\balpha$ the limiting value $\balpha_0 = (1, 0)$ which we have already 
noted to be a rotational fixed point.  Recalling here that $\balpha_0$
was the arbitrary value of $\balpha$ that we had chosen to construct our 
Yukawa term (\ref{Yukawas}), we now realize what it represents, namely 
the high scale limit of $\balpha$, i.e.\ when $\mu \longrightarrow \infty$.
Conversely, for decreasing $\mu$, $\balpha$ will rotate in the 
counterclockwise 
direction reaching eventually at $\mu = 0$ the fixed point at 
$\alpha = \pi/2$, i.e.\ $\balpha = (0, 1)$.  

These results are easily generalized.  For example, had we started with 
an $\balpha$ pointing in the 4th quadrant, i.e.\ $- \pi/2 < \alpha < 0$, 
then $\balpha$ will rotate clockwise for increasing $\mu$, so that 
$\balpha = (1, 0)$ will remain the high scale fixed point, but $\balpha$ 
will reach eventually instead at $\mu =0$ the fixed point at 
$\alpha = - \pi/2$ or $\balpha = (0, -1)$.  Furthermore, given the 
intrinsic $\widetilde{su}(2)$ invariance of the system, it is clear that 
the choice of the starting point $\balpha_0$ has no particular meaning.  
One could have started with any other choice for $\balpha_0$ to construct 
the Yukawa coupling, which $\balpha_0$ will then play the role of the 
high scale fixed point, and $\balpha$ will rotate for decreasing $\mu$, 
in either direction depending on the initial condition, for a quarter 
circle till it reaches the low scale fixed point at $\mu =0$.  

The same analysis can be carried out for the actual 3-G case using the
Higgs states listed in (\ref{3GV}).  This can and has been done at present
to leading order in $R$ and for the special class of vacua defined by $A$
of (\ref{Aspecial}).  Again, as in the 2-G case, one finds that all the
``real'' states $|v_i \rangle, |w_i^R \rangle$ are orthogonal, except for 
$|v_1 \rangle$ which is parallel, to $|v_0 \rangle$, so that their total
contribution to the governing vector is just: 
\begin{equation}
\sum_i \langle v_0|v_i \rangle |v_i \rangle
   + \sum_i \langle v_0|w_i^R \rangle |w_i^R \rangle   
= \sum_i \langle v_i|v_0 \rangle |v_i \rangle
   + \sum_i \langle w_i^R|v_0 \rangle |w_i^R \rangle
= |v_0 \rangle,
\label{govR3G}
\end{equation}
giving thus no rotation.  On the other hand, one has from the ``imaginary'' 
states:
\begin{equation}  
 \sum_i \langle v_0|w_i^I \rangle |w_i^I \rangle
   = - \sum_i \langle w_i^I|v_0 \rangle |w_i^I \rangle
   = |{\rm gvI} \rangle,
\label{govI3G}
\end{equation}
with
\begin{equation}
|{\rm gvI} \rangle = 
-\frac{\sqrt{3}}{4} R \sin 2 \theta \left( \begin{array}{c}
   \sin \theta\; e^{i \beta_1}\\ \ \\
\frac{\displaystyle \cos \theta \sin \phi}{\displaystyle \sqrt{1 - 
\sin^2 \theta
   \cos^2 \phi}} \;e^{i(\beta_1+\sigma_3)} \\ \ \\
\frac{\displaystyle \cos \phi \cos^2 \theta}
   {\displaystyle \sqrt{1 - \sin^2 \theta \cos^2 \phi}}
  \;e^{-i(\beta_2+\beta_3+\sigma_3)} 
\end{array} \right). 
\label{gvIket}
\end{equation}
Again, this being neither parallel nor orthogonal to $|v_0 \rangle$,
there will in general be rotation, except when $\sin 2 \theta = 0$
where the rotation vanishes and we have again rotational fixed points.

We shall leave the detailed analysis of the rotation for the 3-G case
for later when the calculation has been made more complete.  We note
only that in restricting $A$ only to the form (\ref{Aspecial}), we
have thrown away a phase which could be important for the understanding
of CP-violation in the mixing of quarks and leptons, which has so far
not made an appearance.  However, this is only a matter of algebraic
complications which we need to sort out in the near future.

Given the rather significant role that the rotating fermion mass matrix
will play in what follows, it would seem worthwhile to give here a brief
resum\'e of the manner it arises in the present framework, which is
surprisingly intricate, involving, as it does, the doubled invariance 
introduced by the framon idea in an essential way as well as a delicate
interplay between the electroweak and strong sectors of the theory.
The mass matrix here being of the factorized form (\ref{massmatW}), its 
rotation is encapsulated in the rotation of the vector $\balpha$.  This
last is a vector in $\widetilde{su}(3)$ space and originates in the weak 
framon $\phi_r^{\tilde{a} \tilde{r}}$ of (\ref{phirrtat}), having at first 
nothing to do with the fundamental fermions (\ref{psis}).  It only got 
attached to the left-handed bound fermion states $\chi^{\tilde{a} \tilde{r}}$
and $\chi_a^{\tilde{a} \tilde{r}}$ of (\ref{chiss}) and (\ref{chis}) 
through their framon constituents, and so appear in the fermion mass 
matrix (\ref{massmatW}).  This vector $\balpha$, carrying only the global
$\widetilde{su}(3)$ index, has by itself no gauge interaction to give
it rotation, but by virtue of its coupling to the strong sector via the 
linkage $\nu_2$ term in the framon potential (\ref{VPhivec}), it feels the
effects of strong dynamics.  In passing, we note that this $\nu_2$ term
in the framon potential was not introduced by fiat but is a consequence
of the doubled invariance intrinsic to the framon idea.  Apart from
coupling $\balpha$ to the strong sector, this same $\nu_2$ term also
distorts the strong framon vacuum from its original simple configuration 
of an orthonormal triad to a configuration where deviations from both
orthogonality and normality appear.  The vacuum becomes degenerate under
these distortions with deviations from orthogonality traded off with
deviations from normality.  And the vector $\balpha$ is coupled to the
strong sector in such a way that as the vacuum moves among the degenerate
set, the vector $\balpha$ moves (rotates) with it.  That the vacuum does
indeed move, or that the vector $\balpha$ does indeed rotate, was finally
ascertained by studying the renormalization effects to 1-loop order on
the (hadron) fermion self-energy.  Even in this, the interplay between
the electroweak and strong sectors plays an essential role, for without
the vector $\balpha$ coming from the electroweak sector, one finds that
one cannot even construct a Yukawa term for the (hadron) fermion which
is $\widetilde{su}(3)$ invariant as required.  Indeed, as noted in 
\cite{framevec}, treating just the strong sector by itself, even after 
implementing the framon idea, one still would not obtain any rotation.  
Only treating the electroweak and strong sectors together as in the 
present framework and implementing then the framon structure will one 
find that rotation of the fermion mass matrix results.  It would thus 
seem that the rotation of the fermion mass matrix here is a consequence 
of the present framework produced from its very depths.

\setcounter{equation}{0}

\section{Fermion Mixing and Mass Hierarchy}

In a nutshell, that fermion mixing and mass hierarchy would result from 
a rotating fermion mass matrix can be seen as follows.  A mass matrix 
of the simple form (\ref{massmatW}) of a product of a vector $\balpha$ 
with its hermitian conjugate which depends on the fermion species only 
through its normalization $m_T$ means of course that at any chosen value 
of the scale $\mu$, there is only one eigenvector with non-zero eigenvalue, 
namely $\balpha$ with eigenvalue $m_T$, and that this vector, though not 
its eigenvalue, is the same for all fermion species.  However, the masses 
and state vectors of physical particles are not measured all at the same 
scale.  Indeed, the normal convention is to measure these quantities 
each at the scale equal to the particle mass itself.  Thus, for example, 
the state vectors of the $t$ and $b$ quarks are to be taken respectively 
at $\mu = m_t$ and $\mu = m_b$, and being the heaviest state each in its 
own species, they are to be identified as the eigenstate of $m$ with 
nonzero eigenvalue, namely $\balpha$, but one at $\mu = m_t$ and the 
other at $\mu = m_b$, thus:
\begin{equation}
{\bf v}_t = \balpha(\mu = m_t); \ \ \ {\bf v}_b = \balpha(\mu = m_b).
\label{vtvb}
\end{equation}
Since the scales differ, however, and $\balpha$ rotates with scale, it
follows that ${\bf v}_t$ and ${\bf v}_b$ will no longer be aligned, or
that the CKM matrix element:
\begin{equation} 
V_{tb} = {\bf v}_t^* \cdot {\bf v}_b \neq 1.
\label{Vtb}
\end{equation}
Hence mixing between the $U$ and $D$ states.  Further, the $c$ quark, 
being an independent quantum state to $t$, must have a state vector 
${\bf v}_c$ orthogonal to ${\bf v}_t$.  This means that at the scale 
$\mu = m_t$ where ${\bf v}_t$ is defined as the only eigenstate of $m$
with nonzero eigenvalue, ${\bf v}_c$ must be an eigenstate of $m$ with 
a zero eigenvalue.  But this is not the mass $m_c$ of the $c$ quark 
which is to be measured instead at the scale $\mu = m_c$, at which 
scale the eigenvector $\balpha$ with the nonzero eigenvalue would have 
rotated already to a different direction and no longer orthogonal to 
the state ${\bf v}_c$.  At $\mu = m_c$, therefore, ${\bf v}_c$ can no 
longer be an eigenvector with zero eigenvalue but, having now a component 
in the direction of the massive state $\balpha$, would have acquired a 
nonzero mass, as if by ``leakage'' from the heavy state.  Hence the 
mass hierarchy.   Details for working out mixing matrices and lower
generation masses from a given rotating mass matrix can be found in
\cite{physcons,ckm,phenodsm,genmixdsm} in the context of our ealier
phenomenological model, the DSM, or in a general context, and thus
more lucid language, in a recent note \cite{strongcp}.

That a rotating fermion mass matrix can give a reasonable description of 
the fermion mass hierarchy and of the mixing phenomena including neutrino
oscillations observed in experiment has been demonstrated in a number 
of earlier articles \cite{phenodsm,genmixdsm}.  Indeed, it can even be 
said that the data already give circumstantial evidence for mass matrix 
rotation \cite{cevidsm}.  Further, an explicit model (DSM) was constructed
which was able to reproduce most mass ratios and mixing angles to within 
present experimental errors starting from just a small number of fitted 
parameters.  However, it was noted in \cite{phenodsm}, based on the 
experience gained from the fits, that the details of the model are not 
so crucial, but that other models with a rotating mass matrix of the 
form (\ref{massmatW}) with rotational fixed points at $\mu = 0 $ and 
$\mu = \infty$, and a few parameters to adjust, may quite likely do as 
well.

Very briefly, this can be seen as follows.  In the picture with a rotating
mass matrix outlined at the beginning of this section, both the masses of
lower generation fermions and the mixing between up and down fermion states 
arise from the rotation.  Hence, the slower the rotation with respect to 
the change of scale, the smaller will also be the resultant effects.  Now
suppose that given a model with the stated properties, one is able to
choose its parameters so as to have the heavy fermions, such as $t$ and
$b$, close to the high energy fixed point at $\mu = \infty$ and the very
light neutrinos close to the low energy fixed point at $\mu = 0$, while the 
remaining fermions with intermediate masses lie somewhere in between where
the rotation is faster, then, apart from the general features of up-down 
mixing and mass hierarchy already noted, the following empirical facts 
will automatically result:
\begin{description}
\item{(i)} $m_c/m_t < m_s/m_b < m_\mu/m_\tau$ by virtue of the relative
proximity of $t, b, \tau$ in that order to the high energy fixed point
so that the ``leakage'' of masses to the lower generation $c, s, \mu$ is
increasing in that order.
\item{(ii)} The mixing between quarks is considerably smaller than that 
between leptons, by virtue of the heavier masses of the quarks
which place them
closer to the high energy fixed point than the leptons.
\item{(iii)} The corner elements of the mixing matrices, i.e.\ $V_{ub}$
and $V_{td}$ in the CKM matrix for quarks, and $U_{e3}$ in the MNS matrix
for leptons, are much smaller than the other elements, by virtue of the
geometrical fact that the corner elements are associated with the torsion
but the others with the curvature of the rotation trajectory traced out by 
$\balpha$ through changing scale \cite{features}. 
\item{(iv)} The element $U_{\mu 3}$ which governs the oscillation of
atmospheric neutrinos is near maximal by virtue of the neutrino's close 
proximity to the low energy fixed point at $\mu = 0$.
\end{description}
\noindent These already encompass most salient features of the fermion mass 
and mixing patterns known experimentally to-date.

Furthermore, a scheme where lower generation fermion masses are obtained
by ``leakage'' via a rotating factorizable mass matrix has the following
added attraction.  Since the mass matrix remains factorizable at all
scales, it has always some zero eigenvalues although the actual fermion
masses are nonzero, which means that the QCD
phase angle $\theta$ can be rotated 
away and the strong CP problem avoided \cite{strongcp}.

Now, the present framework differs from the old DSM considerably in its 
starting premises and in all its structural details.  Nevertheless it has 
surprisingly retained a similarity with DSM as regards precisely those 
properties noted above as relevant for deriving the physical effects in 
mind.  First, it has a degenerate vacuum depending on orientation as DSM 
had, which is the initial condition needed for rotation to take place.  
Then it is found to give quark and lepton mass matrices of the required 
factorized form which rotates as the scale changes, and that this rotation 
has a fixed point at $\mu = 0$ and another at $\mu = \infty$.  It seems 
thus quite probable, judging from past experience, that the present 
framework will give (or can be made to give by appropriate adjustments 
of parameters) similar agreement with experiment as DSM did, although 
whether this will indeed be the case has yet to be demonstrated by 
explicit calculations.

Assuming optimistically for the moment that this will turn out to be the 
case, let us examine what we have gained with the present framework in 
comparison with the old DSM \cite{phenodsm,genmixdsm}.  First, we have 
gained on the theoretical basis.  The DSM was a phenomenological model 
constructed for the explicit purpose of understanding the generations 
phenomenon without too high a demand for theoretical consistency, and some 
of the assumptions made were a little ad hoc.  In the present framework, on 
the other hand, we have chosen to start with just a few main assumptions, 
and then to work systematically through the consequences to their logical 
conclusion.  Thus, while the present framework has the hope to be developed 
into a theory, the DSM would have to remain a phenomenological model.

Secondly, there have been major improvements too on specific points, both 
in principle and in practice.  We list in particular the following, which 
serve also to highlight some intricate features of the present model we 
find attractive.

(a) One outstanding weakness of the old DSM is that the electroweak symmetry
$u(1) \times su(2)$ has never been properly incorporated.  In the present
framework, not only is the electroweak symmetry fully incorporated, but is 
treated according to the same general principles as $su(3)$, the other local  
gauge symmetry.  Thus all the local symmetries $u(1), su(2), su(3)$ are taken 
to remain exact, and both nonabelian symmetries to be confining, and all
are associated through the framon idea with each a ``dual'' global symmetry,
i.e.\ $\tilde{u}(1), \widetilde{su}(2), \widetilde{su}(3)$.  While the global
abelian symmetry $\tilde{u}(1)$ remains exact and gives rise to baryon-lepton
conservation as a bonus, the nonabelian global symmetries are both broken.
The breaking is different for $\widetilde{su}(2)$ and $\widetilde{su}(3)$,
but this difference is not imposed by fiat but arises as a consequence of the 
difference in structure between $su(2)$ and $su(3)$ which allows the framons 
of $su(2)$ to satisfy the stronger unitarity constraint (\ref{su2ortho}) 
than that admissible for $su(3)$ framons.  This forces the ``weak'' framons 
$\bphi^{\tilde{r}},\  
\tilde{r} = 1, 2$ in $su(2)$ to have opposite $u(1)$ charges, hence
breaking by their different $u(1)$ gauge interactions the $\widetilde{su}(2)$ 
invariance.  The ``strong'' framons, on the other hand, have all the same 
$u(1)$ charge.  Hence, in contrast, $u(1)$ gauge interaction does not break 
$\widetilde{su}(3)$ invariance.  What breaks $\widetilde{su}(3)$ invariance 
instead is the linkage term $\nu_2$ in the framon potential via the vector 
$\balpha$ originating in $su(2)$.  One sees thus that it is ultimately the 
difference in basic structure of the 2 symmetries $su(2)$ and $su(3)$, and 
not the manner they are treated, which is responsible for the difference 
in physical outcome between the 2 cases.

(b) A second weakness of the old DSM is that the self-interaction potential
of scalar fields is not unique.  That potential was constructed on the basis
of an assumed permutation symmetry between the frame vector fields, which
though not unreasonable, is essentially phenomenological and ad hoc.  Even
then, the potential there is not the only one which can be constructed with
this symmetry.  In contrast, the framon potential (\ref{VPhi}) of the present
scheme is constructed on the basis of the doubled invariance under both the
original local gauge symmetries $u(1) \times su(2) \times su(3)$ and their
associated global symmetries $\tilde{u}(1) \times \widetilde{su}(2) \times
\widetilde{su}(3)$, an invariance embedded already in the framon idea.  It
is also the only potential that we have been able to construct with the
prescribed invariance up to fourth order for renormalizability.  This means
that under radiative corrections the potential will remain of the same 
form, only changing the values of the 7 coupling parameters which appear
there.  Such will not be the case for the potential in DSM, which under
radiative corrections can acquire new terms not yet included.  Moreover,
this unique framon potential of the present scheme has some very attractive
properties.  It has a part involving the ``weak'' framons only which is 
of the same form as that for the standard electroweak theory, and another
part involving the ``strong'' framons only which is close to the potential
of DSM constructed for its favourable phenomenological properties.  What
is most interesting, however, are the additional $\nu_1, \nu_2$ terms 
linking the ``weak'' and ``strong'' sectors which are automatically
admitted and required by the invariance.  These not only distort the 
``strong'' sector vacuum in such a way as to make it partake even more of 
the phenomenologically favourable properties of the DSM potential, e.g.\ 
to make its vacuum degenerate and amenable to rotation, but also couple 
automatically the ``weak'' and ``strong'' sectors in such a way that this 
rotation originating in the ``strong'' sector is carried over into the 
``weak'' sector where the quarks and leptons of primary interest occur.

(c)  In DSM, there being no requirement of $\widetilde{su}(3)$ invariance, 
the Yukawa term was constructed just by summing over the free $\tilde{a}$ 
indices so as to guarantee at least permutation symmetry, which we have 
always regarded as a somewhat weak assumption.  Nevertheless, it was this
assumption which led to the factorizable fermion mass matrix, a rather
important requirement for the scheme to work.  In the present scheme, one 
has instead the $\widetilde{su}(3)$ invariance which is inherent already 
in the framon idea, and this allowed one to construct seemingly unique 
Yukawa terms for both the weak and strong sectors leading automatically 
to factorizable fermion matrices.  In constructing the Yukawa term for
the strong sector one had to rely on the vector $\balpha_0$ coming from
the weak sector
to maintain 
$\widetilde{su}(3)$ invariance, while for the weak sector 
one had to rely on the
vector $\balpha^{(\pm)}$ coming from the electromagnetic sector to
maintain $\widetilde{su}(2)$ invariance.  Interestingly, it is exactly
this intricate interplay between the various symmetries imposed by the
framon idea on the Yukawa terms and on the framon potential in (b)
that has allowed one  not only to keep nearly all of the DSM's 
phenomenologically
favourable features but also, it seems, to avoid most of its pitfalls,
as will be seen in  the following examples.

(d) In DSM, as explained in (a), the electroweak sector was not properly
incorporated, and no consideration was yet given to the confinement picture
of symmetry breaking as set out in our assumption (B) in the introduction.
In the present language when assumption (B) is adopted, the states we
called quarks and leptons in DSM would now appear as bound states of the
fundamental fermion fields with framons confined via colour $su(3)$.  In
other words, they would be hadrons, not quasi-elementary states in what
we now call the standard model scenario.  Of course, one need not perhaps
insist on the confinement picture, but this would seem now unnatural in
view of its success in other circumstances.  Instead, in the present scheme,
quarks and leptons, appearing now as bound states via the supposedly much 
deeper $su(2)$ confinement, would appear as quasi-elementary under present 
experimental conditions.

(e) There was a mystery in the DSM which was not resolved within its
own context.  In writing down the fermion mass matrix in a form analogous
to (\ref{massmatW}) above, the vector, say $\balpha'$, which appeared 
there originated in the strong sector and could depend in principle, as
functions of the scale $\mu$, on the fermion species, i.e.\  whether $U$,
$D$, charged leptons, or neutrinos.  
However, to our surprise, in fitting data
then \cite{ckm} with these $\balpha'$ as parameters, it turned out that 
the best fits for the first 3 species were accurately identical.  So
much so, indeed, that for all subsequent fits, we have just taken the
same $\balpha'$ for all species.  This gives the, to us, very appealing 
picture of all fermions lying on the same rotation trajectory as depicted 
in Figure 3 of \cite{phenodsm} or Figure 7 of \cite{genmixdsm}, and as 
utilised implicitly in coming to the conclusions (i)---(iv) above.  Why 
the trajectories for all $\balpha'$'s should be the same, however, was
not explained in DSM.  Here, in contrast, as asserted in the preceding 
entry (b), the vector $\balpha$ which appears in the quark and lepton mass 
matrix (\ref{massmatW}) originates in the weak framon and does not depend 
on the fermion species.  The rotation mechanism, moreover, comes
from the strong sector and gets transmitted to the vector $\balpha$ only
through the linkage term $\nu_2$ in the framon potential which contains
no species dependence.  In other words, that all fermion species have the
same rotation trajectory, which was an empirical observation deduced from 
fitting experimental data in DSM, seems now to be an automatic prediction in 
the present framework.  An analogous observation applies to the normalisation
of $\balpha$ which is here a matter of definition, whereas for its parallel
$\balpha'$ in DSM, the normalisation changed as the scale changed and 
caused there some problems.

(f) One unhappy feature of DSM was the existence in the Higgs spectrum of 2 
zero modes \cite{ckm}.  These arose in the old model as a direct consequence 
of the 2-dimensional degeneracy of the vacuum just mentioned, i.e.\ 
in addition 
to the degeneracy connected to the local gauge invariance of the system and 
cannot thus be eaten up by the vector bosons.  These Higgs zero modes were a 
headache, for they carry generation indices but no (up-down) flavour index,
i.e.\ charge, which meant that they could give rise to FCNC effects 
unsuppressed
by a mass.  Although in the cases we have examined \cite{fcnc}, disaster was
avoided by some special features in their couplings, they remained a major
concern for the model.  In the present framework, however, there are no
corresponding zero modes.  Recall the Higgs mass matrix given explicitly
above in (\ref{MHiggs}) for the 2-G model.  There is no zero mode there
although, given the analogous 1-dimensional degeneracy of the vacuum in
the 2-G model, one might expect one zero mode by analogy.  The reason
for this difference between the two schemes is a rather subtle one.  In
both, the degeneracy of the vacuum arises, of course, from a symmetry; not 
the local gauge symmetry, but in DSM from the permutation symmetry mentioned
in (b) and in the present framework from the global $\widetilde{su}(3)$
symmetry.  The difference, however, lies in the fact that the symmetry 
index in the present framework is carried by the framon only in the global 
factor $\balpha$ which is $x$-independent, as seen in (\ref{phirrtat}), while 
in DSM it is carried by the $x$-dependent field itself.  Now a field can 
fluctuate and give rise to Higgs bosons, but not an $x$-independent global 
factor like $\balpha$; hence the difference.  The absence of these zero 
modes in the present model has thus removed a big worry as regards possible 
violations of experimental FCNC bounds.

(g) In DSM, the rotating vector called $\balpha'$ in (e) above which gives
the fermion mass hierarchy and mixing patterns is actually the vector 
${\bf r} = (x, y, z)$ introduced before in (\ref{intxyz}), and this rotates
from the high scale ($\mu = \infty$) fixed point at ${\bf r} = (1, 0, 0)$ 
to the low scale ($\mu = 0$) fixed point at ${\bf r} = \frac{1}{\sqrt{3}}
(1, 1, 1)$ .  To fit experimental data, this ${\bf r}$ has to go pretty much
all the way between these 2 fixed points within the physical range from
$\mu = m_t$ the top-quark mass, to $\mu = m_\nu$ the neutrino
masses, i.e.\ the 14 orders of magnitude or so in energy accessed by present 
experiment.  Embarrassingly, however, the gauge and Higgs boson masses in
the model depend on the values of ${\bf r}$, with some e.g.\ proportional 
to $\sqrt{y^2 + z^2}$, and will thus vary greatly with scale when ${\bf r}$ 
varies over the above range.  This fact puts serious constraints on the
DSM's phenomenological applicability.  In the present scheme, there is no
parallel difficulty.  Although the vector $\balpha$ which appears in the
quark and lepton mass matrices does vary over a similar range as ${\bf r}$
in DSM, which ought to be sufficient for fitting the same experimental data
on fermion mass and mixing patterns, the quantities $x, y, z$ on which the
boson masses depend never change much from one another, with differences
$\Delta$ always less than $R$ which has supposedly a small value.   

Besides, there are bonuses such as an explanation for baryon-lepton number
conservation and the possibility of understanding CP-violation in mixing
which was beyond the older model.  Thus so long as the projection made
before of the present scheme's viability in phenomenological application 
is confirmed, then one would end up with a much stronger framework
both from the theoretical and from the practical standpoint.

\setcounter{equation}{0}

\section{Concluding Remarks}

Let us begin by counting up what seem to have been gained:

\begin{description}

\item{(I)} One has assigned to scalar fields a geometrical significance 
that they previously lacked.

\item{(II)} One has gained thereby a theoretical criterion on what scalar 
fields are to be introduced into a gauge theory and needs no longer rely 
entirely on the dictates of experiment.

\item{(III)} One has answered, or at least bypassed, the question why, of 
the 2 nonabelian symmetries in the Standard Model, one ($su(2)$) is broken, 
but the other ($su(3)$) is confined.  One now asserts instead that all 
local gauge symmetries remain exact, and both nonabelian gauge symmetries 
are confining.  It is only the global symmetries $\widetilde{su}(2)$ and 
$\widetilde{su}(3)$ arising by virtue of the framon idea in (I), which are 
broken.  

\item{(IV)} One has gained a solution to the old puzzle of baryon number
conservation (in its modern form of $B-L$ conservation). This turns out 
here to be the same as conservation of the $\tilde{u}(1)$ charge, again 
a consequence of the framon idea (I), which quarks and leptons, as bound 
states of $su(2)$ confinement, have acquired from framons as one of their 
constituents. 
   
\item{(V)} One has recovered the electroweak theory in its standard form.

\item{(VI)} One has acquired for quarks and leptons, again from framons 
through $su(2)$ confinement, an index referring to the global symmetry
$\widetilde{su}(3)$ which can play the role of generations.  So fermion
generations is ``dual'' to colour as baryon-lepton number is to electric 
charge, or as up-down flavour is to the confining $su(2)$.  

\item{(VII)} One has reproduced, with a structurally very different but
theoretically much more consistent scheme, the essential features of an 
earlier phenomenological model which was successful in explaining the 
mass and mixing patterns of leptons and quarks observed in experiments.

\end{description}

A key note in the theoretical construction leading to 
the above enumerated
gains seems to be that of economy.  Not only has no 
local gauge symmetry larger
than that of the Standard Model been introduced and no extension of the
Standard Model framework been made along popular lines as supersymmetry,
higher space-time dimensions, or extended structures like strings and 
branes, but even within the present scheme itself, any ambiguities which
arise (such as embeddings, representations etc.) have been consistently 
settled by an insistence on minimality, and in almost every case, nature
seems to have agreed.  It is not that the present scheme cannot admit of
extensions by supersymmetry, high space-time dimensions or string-like
or brane-like structures, for we can see no contradiction in principle
of the present scheme with all those, but that for answering the questions 
posed at the beginning that interest us, there seems no need at present 
for such extensions.  When approached as suggested here, the answers 
sought appear to be there already within the Standard Model framework.  

Comparing now the list above to the questions posed at the beginning in 
the Introduction, one sees that that there is now an answer offered for
most.  There is even a bonus, namely a solution offered for the ancient
puzzle why baryon (or baryon-lepton) number is conserved, which was not 
conceived as one of the original aims.  But there remains one question
unanswered and one to which the answer is incomplete.  The latter is why
the distinction seen in nature between the 2 nonabelian symmetries in the 
Standard Model.  Although in the confining picture, as in (III) above, one 
has reduced the question by taking both $su(2)$ and $su(3)$ as confining, 
it still begs the question why one can assume, as one needs to do and as 
't~Hooft and Banks and Rabinovici did, that $su(2)$ confinement is so 
much deeper than $su(3)$ colour confinement as to be completely undetected 
at present.  The other is why, as assumed here, and also in the conventional 
formulation of the Standard Model, that $su(2)$ doublets are left-handed 
but $su(2)$ singlets right-handed.  In place of answers to these questions, 
we can offer at present only some, perhaps slightly wild, speculations 
which we are not as yet in a position to substantiate.   

When the idea that fermion generations may be dual to colour was first
conceived \cite{physcons}, it was actually envisaged that by duality here
one meant the nonabelian generalisation of the electromagnetic 
or Hodge star duality, as proposed, for example, in \cite{dualsymm}, 
although at that stage, nothing concrete was made out of the proposal.    
As the idea is developed in the present paper, however, the ``duality''
between the local gauge symmetries and their global duals is really just
a change of frames, sharing little of the intricacies of the Hodge star.
For instance, the massive vector bosons $\tilde{B}_\mu$ in the electroweak 
theory as treated here and in \cite{tHooft,Banovici} is coupled with the 
same strength $g_2$ as the gauge boson $B_\mu$.  
In contrast, if $B_\mu$ and
$\tilde{B}_\mu$ were really dual in the (generalised) Hodge-star  sense, we
would expect the 2 coupling strengths to be related by a Dirac quantisation
relation, say:
\begin{equation}  
g_2 \tilde{g}_2 = 4 \pi,
\label{Dirac}
\end{equation}
with $g_2$ being the coupling strength for $B_\mu$ and $\tilde{g}_2$ that
for $\tilde{B}_\mu$.  Besides, if the (generalised) Hodge dual to $su(2)$
exists, as \cite{dualsymm} claims, then there is an old result of 't~Hooft 
which says that the dual symmetry will be broken \cite{tHooft}, meaning
presumably that it will be associated with some massive vector bosons which 
will be coupled with strength $\tilde{g}_2$ and not $g_2$.  Can we then be 
sure that the actual massive vector bosons 
$W-Z$ we presently see in experiment 
are the $\tilde{B}_\mu$ considered above, or are they the (generalised)
Hodge duals here considered?  Suppose for the moment that it is the latter
we see, with the experimentally measured coupling strength of $\alpha_2
= \tilde{g}_2^2/4 \pi \sim 0.034$, then by the Dirac quantisation condition
above, the coupling strength $g_2$ of the gauge boson $B_\mu$ would be very 
large, of the order $20$, much larger than the coupling strength of the 
colour gluon, $g_3 \sim 1.2 $.  One could perhaps then understand why 
$su(2)$ confinement is so much deeper than $su(3)$ colour confinement.  
Besides, if $g_2$ is really that large, then the massive vector bosons 
$\tilde{B}_\mu$ considered before whose mass is proportional to $g_2$ will 
be extremely heavy, probably much heavier than the (generalised) Hodge 
duals that we identify as the $W$'s and $Z$.  If so we would have answered 
why they are not seen, and hence the whole question that was posed.  But 
at the moment, we have no clear idea whether this can indeed be the case.

Amusingly, the speculation aired in the preceding paragraph offers also a
possible answer to another question posed before concerning the apparent
lack of influence on physics at our present ``low'' energies from higher
vector boson states which are expected as radial and orbital excitations of 
the $W, Z-\gamma$ complex when the confinement picture of the electroweak
theory is taken to be physical.  This lack is intuitively understandable,
and is confirmed by analysis in \cite{CFJ}, if the excited states are much 
heavier than the ground states $W, Z-\gamma$, but the question remains why
the states $W, Z-\gamma$ should be so much lighter than their excitations,
which seems in some conflict with our experience in the parallel scenario
of colour $su(3)$ confinement.  However, if our assumption above makes 
sense, then the states $W, Z-\gamma$ are not just the lowest in a tower
of radial and orbital excitations as pictured before, but a completely
different object, namely the dual to the gauge bosons via a nonabelian
generalization to Hodge duality.  This latter is a very complicated state,
in a sense containing already a complex mixture of radially and orbitally
excited components and cannot easily admit, presumably, any further simple 
excitations.  What corresponds instead to the ground states in the familiar 
tower of radial and orbital excitations are not then the $W, Z-\gamma$ 
states but the states called $\tilde{B}_\mu$ in the preceding paragraph, 
which, as the argument goes, are already very much higher in mass than 
the $W, Z-\gamma$ bosons we experimentally observe.  And their excitations 
would be even higher.  There is then little wonder that they have no 
observable influence on the ``low'' energy physics accessible to us at 
present. 

Our speculative answer to the other question why $su(2)$ doublets 
are left-handed but singlets right-handed is perhaps little more than just 
wishful thinking.  The point is that in the confinement picture, doublets
have to be confined, and the actual left-handed quarks and leptons we see
are actually compound states made up of fundamental fermions and framons,
while the right-handed are still just fundamental fermions.  Besides, it
is only the left-handed quarks and lepton which interact with the vector
bosons $W^{\pm}, \gamma-Z$.  Now, suppose for some reason that the framons
should carry with them the projection operator $\frac{1}{2}(1 + \gamma_5)$,
then in all the terms in the action we wrote down, only left-handed flavour
doublets and right-handed flavour singlets would occur, and all the vector
bosons, being bound states of $\bphi^{\dagger} \bphi$, will be left-handed. 
But we can see no reason at present why framons as frame vectors should
carry with them the said projection operator.

However, even with these 2 theoretical questions still unanswered, one has
made, it seems, quite some progress towards a protogenic model.  But what
about phenomenology?  Although the present scheme, as shown, has reproduced 
all the essential features of our earlier phenomenological (DSM) model,
and has therefore, as judged by previous experience, a good chance of
reproducing the earlier model's phenomenological successes, this has to 
be confirmed by calculation, which has started but not yet given definite
results.  Optimistically, there might even be a chance of improving on 
the earlier results, given that there may now be a possibility of getting 
CP-violation in the mixing matrices which was not possible for the earlier 
model.

Then there is a host of phenomenological questions to be examined on 
whether the present scheme may give rise to new phenomena violating 
present experimental bounds.  This will be a lengthy process which will 
take time to sort out.  One of the foremost question to examine, we think, 
would be whether the proposed existence of the ``strong'' framons would 
disturb the apparent agreement of standard QCD with experiment in, for 
example, the running with energy of $\alpha_S$, the strong coupling constant.  
A preliminary investigation says no, given that the contribution of the
scalar framons to the $\beta$-function is very small, only $1/8$ of that
from fermions \cite{Grosczek,Cheichtenli}, 
but this has yet to be systematically confirmed.

Again, optimistically, one may ask whether, apart from explaining known
effects and surviving existing bounds, the present scheme gives some new
characteristic predictions testable by foreseeable experiments.  As far
as we can see at present, any quantitative prediction will have to await
the conclusion of the calculations and investigations projected in the 
last two paragraphs.  One outstanding qualitative prediction, however, 
which might be testable already when the LHC comes into operation, is the
existence of internal structures for leptons and quarks which up to now
have appeared point-like.  This prediction, though present already in the 
confinement picture for the standard electroweak theory as an alternative
interpretation, has now become an apparent imperative in the present scheme 
for, as we recall, it is through $su(2)$ confinement that quarks and leptons 
acquire from their framon constituents both their generation index and 
their baryon-lepton number.  In principle, the internal structures of quarks 
and leptons can be detected by deep inelastic scattering experiments as 
the structure of the proton was detected, only at a much deeper level.  
Unfortunately, our meagre understanding of $su(2)$ confinement at present 
is insufficient to predict at what depth the internal structure of quarks 
and leptons are to be detected.  But for all we know, it may be just around 
the corner, and it would seem to pay the experimenter to look out for it as 
soon as the LHC starts to run. 

The prediction of internal structure for quarks and leptons, though
physically highly significant, is insufficiently special or specific to 
distinguish the present scheme from other composite models.  However, 
the same dynamics that led here to the said internal structure can 
manifest itself also in the existence of excited quark and lepton 
states, as already mentioned, which may be experimentally produced if 
the energy is high enough.  And the spectrum of these will depend on the 
details of the scheme.  Indeed, if one takes the confinement picture 
of the electroweak theory as actually physical, then there would be a 
host of other new states formed by $su(2)$ confinement from any singlet 
combinations of the fundamental framon, gauge and fermion fields.  This 
will open up a whole new field of spectroscopy to future investigation 
at sufficiently high energy which is potentially every bit as rich as, 
if not even richer than, hadron spectroscopy.  

Of most exotic interest in this scenario
would perhaps be those states formed from a pair 
of the fundamental fermion fields $\psi(2,1)$ or $\psi(2,3)$, which 
(though bosonic) are analogous to baryons in hadron spectroscopy.  They 
can be grouped into the following 3 types: $\psi(2,1) \psi(2,1)$, 
$\psi(2,3) \psi(2,3)$, and $\psi(2,1) \psi(2,3)$, which, for reasons to 
be made clear, we shall label as dileptons, diquarks and lepto-quarks 
respectively.  Generically, of course, dilepton, diquark, and lepto-quark 
states occur in any scheme where the confinement picture for $su(2)$ is 
taken as physical, as already considered in for example \cite{CFJ}.  But 
in the present framework, there is a new twist, because of the particular 
way that the baryon-lepton number and the generation index are introduced.
We recall that the baryon-lepton number as well as the generation index
in the present scheme are attached not to the fundamental fermion fields 
but to the framon fields, and only got transmitted to the bound quark 
and lepton states via $su(2)$ confinement of the fundamental fermion 
fields with the framons.  We have thus the unusual situation that what 
were called dileptons, diquarks and lepto-quarks above, comprising as 
they did only fundamental fermion states, do not actually carry a baryon 
or lepton number, nor for that matter a generation index.  Nevertheless, 
when they decay, as they presumably will since they are expected to have 
quite high masses, they would do so most probably, in analogy to what 
hadrons do, by creating a framon-antiframon pair which, by combining 
respectively with the 2 constituent fundamental fermion fields, will 
give back the appropriate baryon-lepton numbers and generation indices
to the quarks and leptons in the decay products.  Hence, the first type
will decay into a pair of leptons, the second into a pair of quarks, 
and the third into a lepton and a quark, as the labels of respectively
dileptons, diquarks and lepto-quarks suggest.  However, the spectrum of 
of these states will be different in the present framework from that 
obtained in models of $su(2)$ confinement \cite{CFJ} where the 
baryon-lepton number and generation index are attached instead to the 
fundamental fermion fields.  
Thus, given sufficiently high energy for this new spectroscopy to be 
explored, there should, we think,
 be no great difficulty in distinguishing the 
present framework from the others.  Clearly, a lot of the details of the 
phenomenology remain yet to be worked out, the treatment of which would 
take us way beyond the scope of the present paper, and has to be left to 
be supplied, we hope, elsewhere later.  But the brief discussion above may 
already serve to indicate the richness of this possible new field of 
particle spectroscopy, which can become a prolific hunting ground for 
future experiments, if not yet already at LHC, then one day when high 
enough energies are available. 

Lastly, we note that of the physical ingredients making up the present
framework, the most distinctive are perhaps the strong framon fields 
$\phi_a^{\tilde{a} \tilde{r}}$, at least when taken as elementary as 
they are in this paper.  It would be nice, therefore, if one could 
device a phenomenological handle for their detection.  These are colour 
triplets and can thus, like quarks, manifest themselves in 2 ways, 
either confined into freely propagating colour singlet hadron states, 
or else appearing as jets in hard scattering, but the detection of 
either will not be altogether easy.  For example, by binding together 
a framon-antiframon pair, we would obtain in the $s$-wave the scalar 
states studied in the second half of section 5, or else in the $p$-wave 
some analogous vector states.  But these are hadrons with presumably 
rather high masses, and hence broad widths, and will not be easy either
to detect or to distinguish at first sight from hadrons of the common 
$q \bar{q}$ type.  Nevertheless, we do know a fair amount about these 
states, as the analysis of section 5 shows, and this can be further 
extended, so that with luck and hard work, sufficient distinctive 
features may be identified for their eventual detection.  The same 
remark applies to the problem of detecting the jets in hard scattering 
originating from framons and distinguishing them from the quark jets 
also produced.  Clearly, there is scope for much future phenomenology 
in this direction.

We are much indebted for constant interest and encouragement as well as for
occasional remarks and help in checking certain arguments to Jos\'e Bordes
who has been in fact our collaborator in spirit throughout the several long 
years that this work has occupied us.

\end{document}